\documentclass[acmlarge,screen]{acmart}

\def\BibTeX{{\rm B\kern-.05em{\sc i\kern-.025em b}\kern-.08emT\kern-.1667em\lower.7ex\hbox{E}\kern-.125emX}}
    
\setcopyright{rightsretained}
\acmDOI{10.1145/1122445.1122456}

\usepackage{hyperref}
\hypersetup{
    colorlinks,
    linkcolor={red!50!black},
    citecolor={blue!50!black},
    urlcolor={blue!80!black}
}
\begin{document}

%
\title{Human-Misinformation interaction: Understanding the interdisciplinary approach needed to computationally combat false information}

%
\author{Alireza Karduni}
\email{akarduni@uncc.edu}
\affiliation{%
  \institution{UNC-Charlotte College of Computing and Informatics}
}

%
\renewcommand{\shortauthors}{Alireza Karduni}

%
\begin{abstract}
The prevalence of new technologies and social media has amplified the effects of misinformation on our societies. Thus, it is necessary to create computational tools to mitigate their effects effectively. This study aims to provide a critical overview of computational approaches concerned with combating misinformation. To this aim, I offer an overview of scholarly definitions of misinformation. I adopt a framework for studying misinformation that suggests paying attention to the \textit{source}, \textit{content}, and \textit{consumers} as the three main elements involved in the process of misinformation and I provide an overview of literature from disciplines of psychology, media studies, and cognitive sciences that deal with each of these elements. Using the framework, I overview the existing computational methods that deal with 1) misinformation detection and fact-checking using Content 2) Identifying untrustworthy Sources and social bots, and 3) Consumer-facing tools and methods aiming to make humans resilient to misinformation. I find that the vast majority of works in computer science and information technology is concerned with the crucial tasks of detection and verification of content and sources of misinformation.
Moreover, I find that computational research focusing on Consumers of Misinformation in Human-Computer Interaction (HCI) and related fields are very sparse and often do not deal with the subtleties of this process. The majority of existing interfaces and systems are less concerned with the usability of the tools rather than the robustness and accuracy of the detection methods. Using this survey, I call for an interdisciplinary approach towards human-misinformation interaction that focuses on building methods and tools that robustly deal with such complex psychological/social phenomena.
\end{abstract}

%

%

%
\maketitle

\section*{Introduction}
Throughout history, misinformation, or information that is false, has been used intentionally to manipulate people's opinions and beliefs \cite{wardle2017information}. Technological advancements have been a crucial element in the development of misinformation. Its earliest traces can be traced back to societies with the earliest writing systems through which rulers would falsify written record to glorify themselves and demean enemies \cite{marcus1992mesoamerican, taylor2013munitions}. The effects of information manipulation for political or economic gain only increased by the invention of new technologies such as print, press, and later in the 20th century by the explosive utilization of mass media such as television and the radio \cite{taylor2013munitions, herman2010manufacturing, arsenault2006conquering}. The birth of the Internet, Information Communication Technologies (ICTs), and social media drastically increased the rate and means of information production, curation, and sharing. 

Consequently, this massive amounts of misinformation affect a large number of people on a global scale \cite{lazer2018science, howell2013digital, newman2017reuters, allcott2017social}. Prevalence of false information in societies has the potential of making democracies ungovernable \cite{benkler2018network}. As the advancements in computation and communication technologies have played a significant role in the growth of misinformation and its threats to our societies, we now face a dire need to develop new technological efforts and tools for effectively battling this growing problem.  

The issue of misinformation today is more complex and multi-faceted than ever. Individuals and organizations can now easily create and disseminate information on social media. Numerous agents with malicious or non-malicious intents participate in creating a phenomenon often labeled as ``fake news'' \cite{wardle2017information, lazer2018science}. Furthermore, sources of misinformation can use automated/semi-automated bots on social media platforms to rapidly spread misinformation \cite{ferrara2016rise, shao2017spread}. Consumers are likely to believe and share without rigorous fact-checking. Cognitive scientists have highlighted various factors such as prior exposure to news\cite{pennycook2018prior}, selective exposure, and confirmation bias causing audiences to believe misinformation \cite{lazer2018science}. News outlets take advantage of these psychological factors and introduce slants, falsities, or political biases into the content of their news \cite{arsenault2006conquering, adams1986whose}.

Additionally, social media platforms tend to introduce algorithms to curate information that has been shown to create filter bubbles or echo chambers through which audiences are less likely to be exposed to news they do not agree with \cite{pariser2011filter,bakshy2015exposure}. There are many terms used by scholars, journalists, and politicians that refer to the accuracy and intentions of information and media including propaganda, disinformation, misinformation and fake news. However, there are no agreed-upon definitions for these types of information manipulations \cite{wardle2017information, tandoc2018defining, jack2017lexicon}. These complexities in sources, types, means of production, and definitions of misinformation along with different psychological and social factors highlight the need for a comprehensive and multi-disciplinary approach towards preventing and intervening misinformation. 

Even though the issue of misinformation and combating it is at the forefront of many political, journalistic, and scholarly discussions, the landscape of attempts on computationally combating misinformation is still inadequate. Most of the existing approaches fall under Automatic Fact Checking (AFC) and detection of misinformation that include automatic or semi-automatic identification, verification, and correction of misinformation. To date, the effectiveness of these methods without human supervision remains very limited \cite{graves2018understanding}. Furthermore, correcting information does not necessarily result in a change in belief \cite{nyhan2010corrections,flynn2017nature,mele2017combating}; while repeating misinformation even for fact-checking might prove to be counterproductive \cite{swire2017role}. Lazer et al. suggest the necessity of a comprehensive strategy on education, empowering individuals, along with a collaborative approach between industry and academia on battling misinformation \cite{lazer2018science}. 

The structure of this survey is as follows: in section \ref{defining} I offer an overview of different terms and definitions related to misinformation. In section \ref{elements-framework}, I describe my adopted framework for studying misinformation which calls for broadening our focus from fact-checking the content of misinformation to include interactions between content, sources, and consumers of misinformation. In section \ref{elements-relationship}, I overview related literature from psychology, cognitive and social sciences categorized by each element. In section \ref{computation}, I survey existing misinformation-related literature in computer science related to content (section \ref{content}), sources (section \ref{source}), and consumers (section \ref{Consumer}). Finally, in section \ref{discussion}, I discuss the lacking areas of the current literature on computation and misinformation and propose an agenda for future interdisciplinary research.

\section{A framework for studying misinformation\label{framework}}

In this section, I propose a framework for studying and categorizing misinformation. I do so by first describing the different definitions needed to describe the misinformation phenomena better. These definitions help us understand how we can categorize misinformation. Then, I describe the different elements involved in the production and consumption of misinformation including producers, the content of misinformation, and the audience. By combining different definitions and recognizing the elements involved in misinformation, we will be able to critically analyze the existing literature involving computation and misinformation.

\subsection{Defining Misinformation\label{defining}}

After United State's presidential election at 2016, the term ``fake news'' has become a topic of interest in many journalistic and political circles around the globe \cite{Trump’s‘26:online,Wielding48:online,wardle2017information}. Fake news has been defined as ``deliberately constructed lies, in the form of news articles, meant to mislead the public'' \cite{Whythete18:online}. Many scholars from different disciplines have elected to use this term as they discuss the problem of misinformation \cite{lazer2018science,allcott2017social,wang2017liar,klein2017fake}. A survey of 34 scholarly articles using the term fake news showed that it is used to describe a wide range of concepts including news satire, news parody, news fabrication, photo manipulation, advertising in the guise of news reports, and propaganda. The authors conclude that the common feature between all mentions of ``fake news'' is that they all "appropriate the look and feel of real news; from how websites look; to how articles are written; to how photos include attributions." \cite{tandoc2018defining}.

Other scholars suggest refraining from using the term fake news; as it generally fails in correctly and accurately describing the complexities of misinformation \cite{wardle2017information, jack2017lexicon, Fakenews76:online, Whythete18:online, Stopsayi1:online}. Starbird focuses on a dichotomy of "alternative" and "mainstream" media outlets. She utilizes conspiracy theories and alternative facts as the term to describe misinformation and "fake news" \cite{starbird2017examining}. However, scholars such as Castells and Chomsky have shown various ways mainstream media utilize misinformation to influence peoples' opinions \cite{herman2010manufacturing,arsenault2006conquering}. Caroline Jack from Data and Society calls for more accurate terminology for discussing "problematic information" because each word might infer assumptions about the producer, the type of message, and the persons receiving the information. The report emphasizes the importance of paying attention to these factors as they greatly affect the strategies needed for combating them. She differentiates between misinformation and disinformation by defining misinformation as ``information whose inaccuracy is unintentional'' while disinformation describes ``purposefully falsifying information''. She also highlights the difficulties in differentiating between publicity and propaganda. While both are geared towards influencing audiences, propaganda is often referred to as attempts to deliberately manipulate or deceive people. Moreover, Jackson describes a third category which differs from disinformation or propaganda. While these concepts mostly aim to gain support for different beliefs or idea, some events, described by the term ``gaslighting''  use falsified information to create uncertainty and tension in various societies. \cite{jack2017lexicon}. 

\begin{figure}[t!]
  \caption{A reconstructed diagram describing the categorization of false information by Wardell and Derakhshan\cite{wardle2017information}.}
  \label{categoriesVen}
  \includegraphics[width=0.75\textwidth]{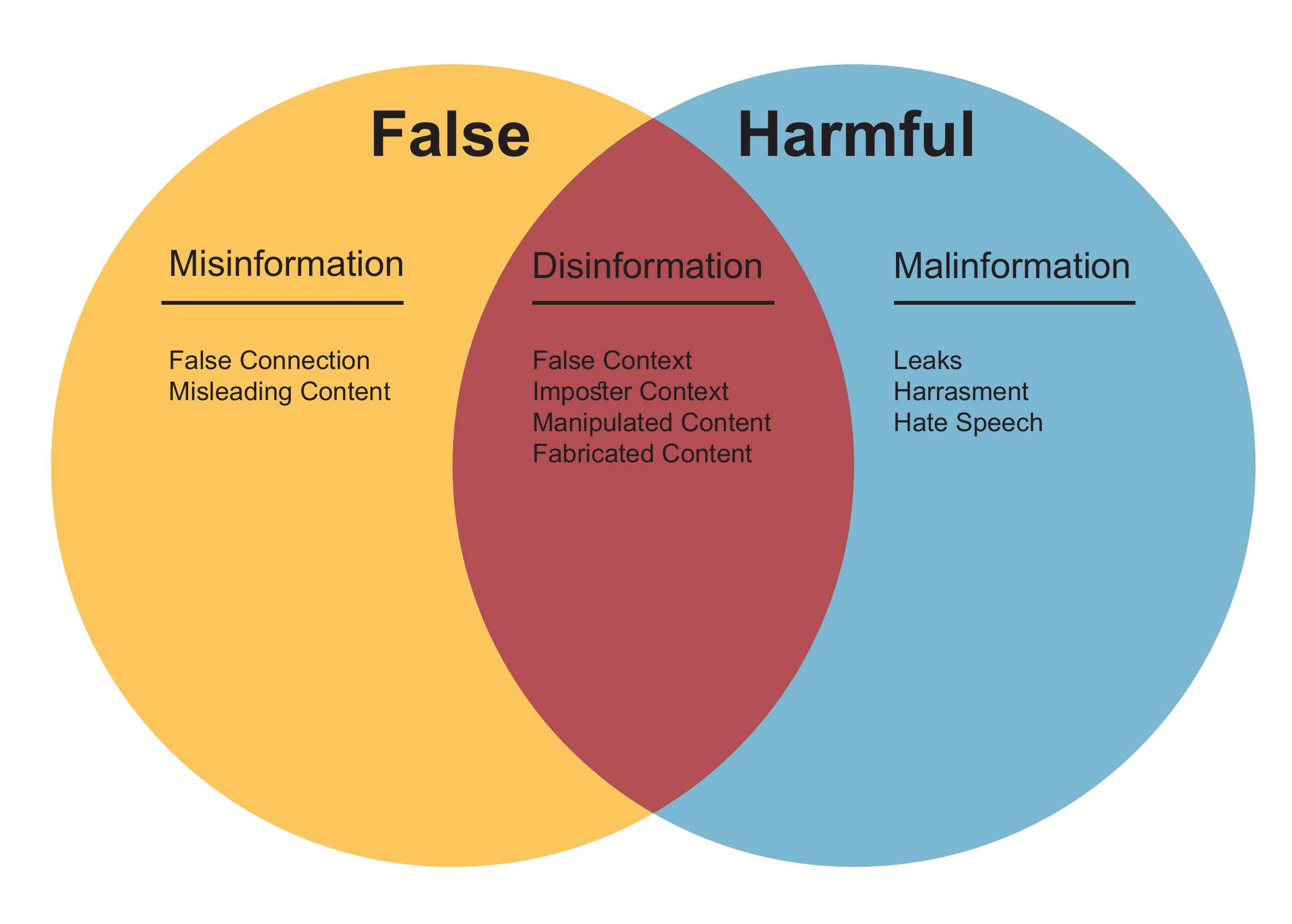}
\end{figure}

A report on "Information Disorder" by Wardell and Derakhshan produced for the Council of Europe, categorizes problematic information into three groups by asking two questions: Whether a piece of information is harmful or not and whether it is false. These categories are Misinformation, Disinformation, and Malinformation \cite{wardle2017information}. Misinformation is information that is false but is not produced with an intent to harm. It is created when journalists or individuals share misinterpretation about an event or a rumor without realizing the information is not accurate. For example, after the Bombing at the Ariana Grande concert in Manchester, images of several individuals were tweeted by multiple agencies as missing. Several of those individuals, in reality, had nothing to do with the bombings \cite{HereIsAl84:online}. Disinformation, on the other hand, is false information that is created with the intention to harm groups or individuals. A well-known example of Disinformation was a conspiracy known as Pizzagate created by Alex Jones of Infowars that Hillary Clinton had sexually abused children in satanic rituals \cite{Pizzagat59:online}. Malinformation is information that is not false but is released, often illegally, with the intent to harm. Two examples of such malinformation are the release of personal emails before two elections in the United States from Hillary Clinton, which is believed to have affected the results of the presidential election, and Emanuel Macron, which was not effective in its intent \cite{WhytheMa1:online}. Figure \ref{categoriesVen} shows the three categories as defined by Wardell and Derakhshan.
\cite{}
As noted by many scholars whose work involves misinformation, fake news is not an adequate terminology for describing the complexity of the problem. For consistency with these suggestions, I define misinformation as \textit{information that is false, with or without the intent to harm or manipulate consumers}. My definitions corresponds to the combined definition of Misinformation and Disinformation offered by Wardell et al. \cite{wardle2017information}. In the next section, I describe the elements involved in the production of misinformation that serve as a framework for studying related computational methods addressing this phenomenon.

\subsection{Elements involved in the production of misinformation\label{elements-framework}}

The defining factor in the study of misinformation is its relationship to ``facts'' and ``truth''. Most technology companies and news outlets approach misinformation by providing rigorous analysis and ``fact-checking'' of news. However, Lazer et al. suggest that communicating the results of fact-checking by itself is not enough and in some cases can prove to be counterproductive \cite{lazer2018science}. One of the biggest challenges in addressing the problem of misinformation is the multiplicity of elements involved in the process of production and consumption of misinformation. Scholars from multiple disciplines have highlighted these elements. 

In an analysis of the social production of misinformation in the United States during the Iraq War in 2004, Arsenault and Castells describe a complex model of misinformation that includes media organizations, political actors, general psychological climate, and the mental frames of the audience. One of the interesting factors they discuss is misinformation producers' usage of specific language using framing, biases, slants, and metaphors. The authors discuss in detail, how these elements synthesize and result in social misperceptions \cite{arsenault2006conquering}. The significance of their study is how they describe a model of misinformation that involves the producers of misinformation, their agenda, their usage of emotions and language, and the audiences emotional and mental state. In a study of "fake news" in the US presidential election of 2016, Bakir and McStay argue that we should see the current social media driven landscape of misinformation in light of the systematic, political and commercial efforts in liberal democracies to influence opinions and beliefs of populations through propaganda. They count five main elements as crucial in the current landscape of misinformation: "The financial decline of legacy news; the news cycle's increasing immediacy; the rapid circulation of misinformation and disinformation via user-generated content and propagandists; the increasingly emotionalized nature of online discourse; and the growing number of people financially capitalizing on algorithms used by social media platforms and internet search engines" \cite{bakir2018fake}. They also recognize the multiplicity of elements and the importance of observing the economic benefits and intentions of sources of misinformation, as well as the emotional state of the audience.

Vargo et al. study the agenda-setting power of "fake news" and consider partisan media, "fake news" media, and fact-checkers as three separate entities that influence people's opinions consequently have power in setting domestic and international policy agenda \cite{vargo2018agenda}. Lazer et al. emphasize the importance of focusing not only on the message but also paying attention to the source of misinformation as well as taking an educational approach to address the consumers of misinformation \cite{lazer2018science, mele2017combating}. Wardell and Derakhshan, describe three essential elements including Agent ( or creator ) of misinformation, the message, and the interpreter. For each of these elements, the authors describe multiple vital factors to consider including the intent and type of agent, the emotional content and type of the message,  and the mental state of the interpreter\cite{wardle2017information}.

In the report by Wardel and Derakhshan curated for the council for Europe, the authors define three main elements in the process of ``information disorder``: 1) The agent or the actors who initially distribute the message, 2) the message which is the content of misinformation and encompasses text, images, and other types of media, 3) and the interpreter who is the person consuming the message. Inspired by these three elements while taking into account other scholarly definitions \cite{lazer2018science,tandoc2018defining,arsenault2006conquering,mele2017combating}, I adopt three main elements of misinformation for conducting critical analysis on means of computationally battling misinformation: source, content, and consumers of misinformation. The reason I chose this terminology is that it more accurately represents different elements of misinformation. In contrast to Wardel and Derakhshan who describe the agent as actors who initially "create and produce and distribute the message" \cite{wardle2017information}; I use the term source to more broadly describe any source including bots and redistributors of misinformation that have not initially created a message. I also use content to more broadly refer to the textual, visual and multimedia content of misinformation. Finally, I use consumers as a broader definition to include groups or individuals who might be the target of misinformation.  

\begin{itemize}
    \item \textbf{Source:} The outlet through which misinformation is being consumed. these outlets can include:
    \begin{itemize}
        \item professional News accounts or agencies
        \item Journalists, bloggers, social media personalities, or news aggregates
        \item Bots that disseminate information whether detected or undetected.
        \item Other individuals who share misinformation
    \end{itemize}
    \item \textbf{Content:} the content of misinformation that is being distributed. The content can include:
    \begin{itemize}
        \item textual messages on social media
        \item news online from news agencies and blog
        \item visual information including images and videos
    \end{itemize}
        \item \textbf{Consumer:} Individuals or groups who are exposed to or effected by misinformation who make various decisions in regards to misinformation:
            \begin{itemize}
        \item they make many decisions when exposed to these content: to trust the content, to trust the source, and to share and propagate the content.
    \end{itemize}
\end{itemize}

Using these elements as a framework for analyzing misinformation allows us to systematically study misinformation from creation or dissemination by sources to consumption by the audience. Through inquiries about the relationship between these elements, we can raise important questions about misinformation. By looking at the relationship between sources and contents of misinformation, we can start to question the means, and methods sources use to create misinformation. Also, by looking at the relationship between consumers and the content, we can explore the psychological and social factors of why individuals believe in misinformation. In the next sections, I explore literature related to each of these elements.\break

\subsection{Elements of misinformation and their relationships\label{elements-relationship}}

\subsubsection{content-truth relationship: facts, fact-checking, and truth}

The relationship between misinformation and facts is likely the most prominent aspect of fighting misinformation. Fact-checking is the primary means of dealing with this relationship and is the main approach by social media platforms, technology companies, media outlets, and third-party organizations that specialize in this task. Social Media platforms take part in the process by taking various measures such as employing third party fact-checkers and developing new technologies to automatically detect fake news \cite{Thisisho46:online,Facebook81:online, GoogleFa66:online}. Many third-party organizations exist that specialize in flagging misinformation content and sources and communicating the results with consumers \cite{snopes.com_2018,FactChec59:online,LatestEm20:online,Factchec51:online}. These organizations provide a variety of information such as different scales of rating and labeling news pieces. They cover political statements, claims, TV ads, and articles. These organizations tag news pieces as suspicious, completely false, misleading, or out of context. There are also numerous fact-checking attempts that use crowd-sourcing as their primary means of battling misinformation\cite{Fiskkitc60:online, FactChec34:online}. These systems are still early in development, and their efficacy is yet to be examined. Even though fact-checking misinformation is a critical task, its effectiveness in battling misinformation has been questioned \cite{lazer2018science,lazer2017combating, figueira2017current}. 

\subsubsection{source-consumer relationship: sources' intent}

When studying Misinformation from the perspective of sources, the first question that comes to mind is the intent of source or why the source is delivering misinformation to consumers. Lazer et al. argue that understanding the intent of sources is a primary factor in understanding misinformation and should be given prevalence over the factuality of a single news story \cite{lazer2018science}. Different motivations and intents have been identified about sources of misinformation. These include financial, economic, and advertisement \cite{bakir2018fake, arsenault2006conquering, bourgonje2017clickbait,herman2010manufacturing,tamibini2017advertising}; ideological or partisan\cite{allcott2017social,tambini2017fake}; political and power \cite{arsenault2006conquering, tambini2017fake,tandoc2018defining}, and satire and parody\cite{day2012live, baym2013news, wardle2017information}. Based on these intents, sources take different strategies of content propagation and manipulation to influence Consumers.

\subsubsection{Source-content relationship: means of propagation, verbal and visual frames, strategies}

The means of which sources propagate misinformation can be an indicator of its intentions. Besides prominent misinformation sources with strong editorial staff, many take extensive use of social bots. Social bots are emerging phenomena that act as sources of misinformation \cite{bessi2016social}. They impersonate real sources of misinformation and are often automatic or semi-automatic. They have the power to rapidly propagate misinformation and pollute the information space \cite{ferrara2016rise}. Social bots are active in the early stages in the life cycle of misinformation and also interact with high profile accounts on social media \cite{shao2017spread}. Social bots tend to behave similarly to real individuals and sources on social media by liking, sharing, and commenting on other sources of news \cite{lazer2018science}. By amplifying the spread of misinformation, social bots take part in the creation of echo-chambers and activating consumers' cognitive biases \cite{lazer2017combating}.

Sources do not treat information uniformly, and they take different manipulation strategies and framing techniques to produce believable stories. News media that produce misinformation often share information in a biased manner. They focus on partisan topics and issues, as well as inflammatory topics, emotional content, specific moral foundations, or specific geographies and political figures. They often focus on topics from online partisan news media and also have the power to set the issue agenda for those accounts\cite{vargo2017agenda}. Moreover, these sources tend to disseminate news about topics that are important to specific populations. These focuses include biases towards extreme ideologies, political parties and figures, and inflammatory issues\cite{spohr2017fake, arsenault2006conquering,allcott2017social}. These topics are chosen based on the psychological climate of audiences often focusing on fear-mongering, amplifying anger and outrage in consumers.\cite{arsenault2006conquering,bakir2018fake}. Sources of Misinformation, often focus on specific moral visions and foundations of different groups, to take advantage of their mental framing.\cite{entman2010media,arsenault2006conquering,lakoff2010moral}. Moreover, they tend to cover news related to only a subset of geographic places, as well as political figures \cite{adams1986whose,allcott2017social}. 
 
 Misinformation sources take extensive use of visual information and images to mislead consumers. They use images to convey different biases not necessarily detectable in textual content\cite{frenkel_2017}. Images contain implicit visual propositioning that can communicate various ideological or stereotypical messages that might receive greater resistance when are put in words. Thus viewers and consumers are more likely to be unaware of the implicit biases and frames in visual content. \cite{abraham2006framing, messaris2001role}. This powerful tool has been used in various contexts including the negative portrayal of different political candidates from different parties\cite{grabe2009image}, out-of-context usage and altering of images\cite{mallonee_2017}, or by merging multiple images to change the implicit message \cite{carlson2009reality}. Images of people, either prominent or not, has significant effects on consumers' opinions. It has been shown that altering images can produce significant differences in people's assessments of a political figure \cite{coleman2010framing}.
 
\subsubsection{Consumer-content relationship: social factors and cognitive processes}

Sources produce misinformation with different intents and using different strategies. However, the reason they are often successful in communicating their stories is only partially due to their approach. Humans are affected by multiple social, cognitive, and psychological processes that make them prone to believe and further share misinformation. 

Various social phenomena have been identified to influence consumers' opinions and decisions toward misinformation. Allcott and colleagues showed that propagation and belief of misinformation were influenced by ideological polarization of consumers \cite{allcott2017social}. Moreover, Pennycook and Rand's research on individuals receptivity towards fake news showed that conservative, right-leaning individuals were more likely to believe misinformation \cite{pennycook2018falls}. This polarization can be a direct result of ``echo chambers`` influenced by the ``filter bubble'' phenomenon where algorithms by technology companies shape the information consumers are exposed to \cite{spohr2017fake}. Even though some studies show that there is no reliable evidence for the existence of filter bubbles \cite{zuiderveen2016should} we can find many examples of echo chambers on various social media platforms \cite{sunstein2001echo,flaxman2016filter,barbera2015tweeting}.
Moreover, individuals are more likely to be affected by misinformation when in a social setting\cite{gabbert2004say}. Moreover, a study on students' tendencies to share misinformation showed that many social reasons such as ``sharing eye-catching messages'' or ``interacting with friends'' ranked high as causes of sharing misinformation \cite{chen2015students}. Overall, it has been noted that we are more likely to believe a piece of information if our social circles also accept them \cite{sloman2018knowledge, lazer2017combating, eckles2017bias}.

Many causes of why consumers believe misinformation, has been attributed to psychological and cognitive factors. These factors include different cognitive biases as well as different degrees of analytical thinking. Kahneman and Tversky describe a form of cognitive bias popularized as Availability Bias in which a person evaluates a probability based on how easily relevant information come to mind \cite{tversky1973availability, spohr2017fake}. This form of bias highlights how access and exposure to specific information can have an impact on opinions and decisions. On a study on causes of misinformation, Pennycook and colleagues showed that prior exposure indeed significantly affects consumers perception of the accuracy of misinformation \cite{pennycook2017prior}. Another aspect of why consumers believe misinformation is selective exposure which can be described as consumers' tendency to believe information that aligns well with their views and beliefs and also avoid information that is against their pre-conceived notions \cite{spohr2017fake, lazer2017combating, karduni2018can}. One of the main causes of this selective exposure is known to be confirmation bias, or our tendency to privilege evidence that confirms our existing hypothesis over all possible hypotheses  \cite{mynatt1977confirmation, lazer2017combating}. Consumers' emotional proximity to different topics and events also increases their susceptibility to believing misinformation \cite{huang2015connected}. In contrast to consumer's cognitive and emotional state, the ability to perform analytically thinking and problem-solving has been shown to increase our resiliency to misinformation \cite{pennycook2018cognitive,pennycook2018falls}.

\begin{figure}[t!]
  \caption{Elements of misinformation and their relations.}
  \label{elements}
  \includegraphics[width=0.9\textwidth]{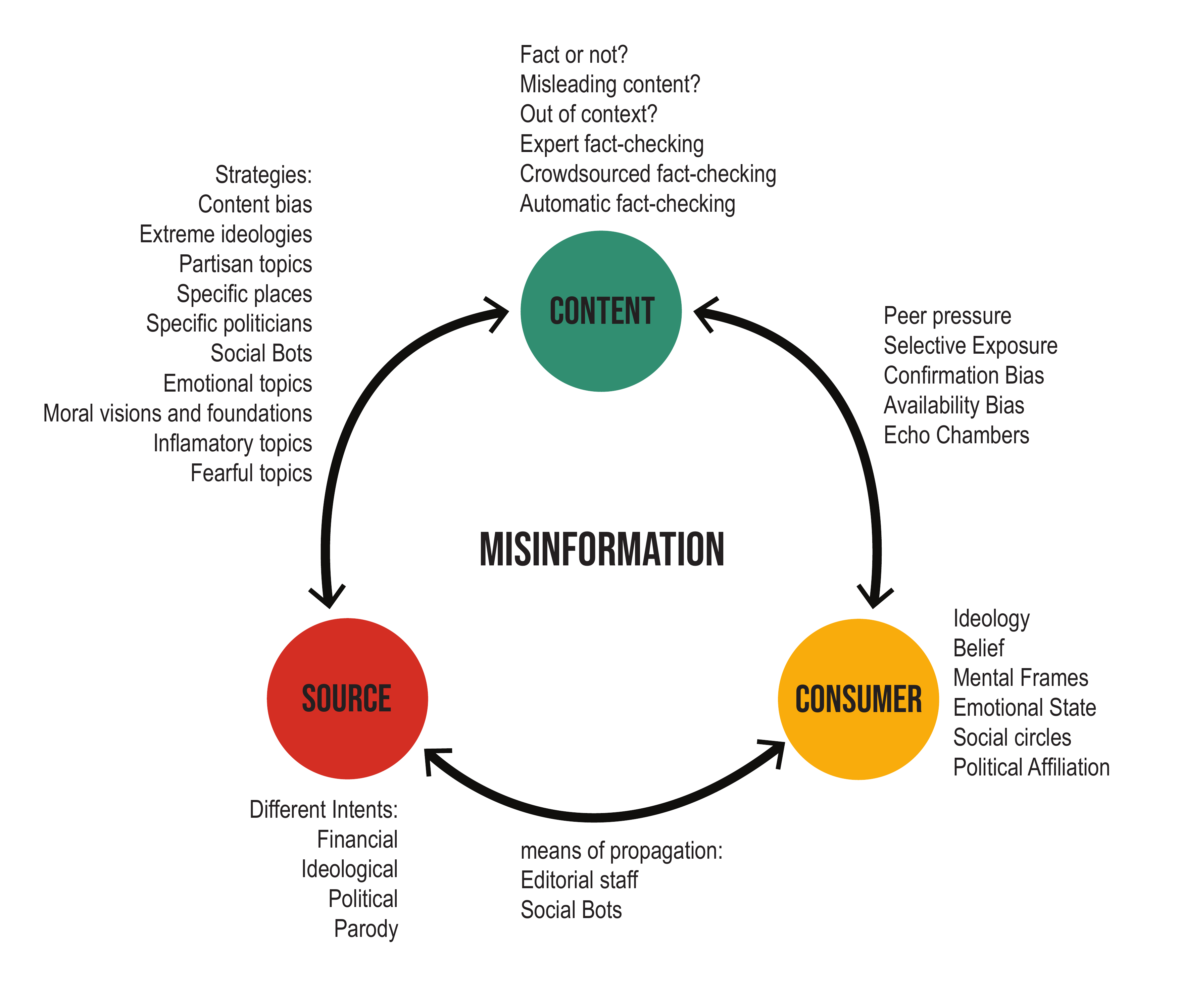}
\end{figure}

Focusing on elements involved in the production and consumption of misinformation and their relationships provides us with a systematic framework derived from multiple disciplines to study how we can use computation to battle misinformation (see figure  \ref{elements}). In the next section, I provide a summary of the state-of-the-art literature in computation related fields that focus on the problem of fake news and misinformation. section \ref{elements-framework} will serve as a framework for categorizing existing works on misinformation and highlighting any existing gaps in our current approaches.

\section{Misinformation and computation: current approaches \label{computation}}

In this section, I provide an overview of the current scholarly studies on misinformation within the fields related to Computer Science and Information Technology. Each section follows the categorization based on the framework and elements discussed in the previous section. It is important to note that many of the mentioned methods focus on multiple elements of misinformation. However, the categorization is based on whether the primary focus of the scholarly work was on either of the elements. In the coming section, I first discuss efforts on Automatic Fact-checking as well as fake news detection. Next, I provide research that focuses on sources of misinformation and their relationship to misinformation. Finally, I focus on research that focuses on the human aspects of misinformation and computation. 

\subsection{Content: Automatic Fact-Checking and Misinformation detection\label{content}}

One of the main topics in the mainstream media in regards to misinformation is fact-checking the veracity and credibility of news articles. The report for the Reuters Institute for the study of journalism defines Automatic Fact-Checking (AFC) as using technologies to "deciding the truth of public claims and separating legitimate views from misinformation." This task, also dubbed as deception detection is defined as the ``prediction of the chances of a particular news article (e.g., news report, editorial, expose) being intentionally deceptive ( fake, fabricated, staged news, or a hoax)'' \cite{rubin2015deception}. In the report, three main elements are defined for Automated Fact-Checking: 1) Identification which involves monitoring news media and sources, identifying factual statements, and prioritizing claims to check; 2) verification which involves checking with other existing fact-checks, checking against authoritative sources, and unstructured credibility scoring; and 3) correction which involves flagging repeated falsehoods, providing contextual data, and publishing new fact-checks \cite{FACTSHEE77:online}. Several systems exist that try to address one or more from these categories.

A unique example of a system that provides claim detection, verification, and correction is ClaimBuster which combines a combination of Natural Language Processing and Supervised Learning to classify and score whether sentences are "check-worthy" or not \cite{hassan2017toward,hassan2017claimbuster}. They classify claims by presidential candidates into Non-Factual Sentences, Unimportant Factual Sentences, and Check-worthy Factual Sentences. The system produces a score that reflects the extent to which a sentence belongs to the check-worthy group. They extract features from text and use Random Forest to find the most discriminating ones. They use an SVM model combined with words, Part of Speech tags and Entity extractions and achieve a precision of 72\% and a recall of 67\% \cite{hassan2017toward}. Later iterations of the System, assesses check-worthy claims with databases of third-party fact-checks based on similarity. Moreover, it provides supporting and debunking evidence from knowledge bases and web \cite{hassan2017claimbuster}. Other attempts in fact-checking generally use external sources to check specific claims or news pieces\cite{shu2017fake}. They use either open web sources using statistical scoring \cite{magdy2010web} or knowledge graphs that can act as ground-truth for misinformation claims \cite{dong2014knowledge}. 

Other approaches to misinformation detection deviate from fact-checking and focus on categorizing news based on features from their content. Numerous research projects have used writing styles and language features visible in the text. A group of scholarly work uses a ``bag of words'' representation to analyze and distinguish misinformation\cite{rubin2016fake}.These methods include lexicon based dictionaries of moral foundations and subjectivity \cite{volkova2017separating,pennebaker2015development,vrij2007cues} and location-based words \cite{ott2011finding}. \cite{shu2017fake}. Other works have used more complex syntactic analysis such as the concept of deep syntax \cite{perlmutter1968deep} to detect deception with high accuracy in multiple datasets \cite{feng2012syntactic}. Another approach to using linguistic features of news content is by building classifiers based on various language features. Oraby et al. compare an automated bootstrapping method of extracting discriminating features from annotated ``fact'' vs. ``feeling'' arguments with a Naive Bayes classifier and show that both methods perform well in discriminating these features \cite{oraby2017and}. 

Potthast et al. have used a modified version of unmasking, a method to determine whether to articles are from a single author or not \cite{potthast2017stylometric}. The method iteratively removes most distinguishing features from two articles and observes the rates of which cross-validation accuracy drops \cite{koppel2004authorship}. The authors show that various stylistic features can distinguish real news content from hyper-partisan media, satire, and fake news. Bourgonje, Schneider, and Rehm approach the problem of misinformation by focusing on Headlines \cite{bourgonje2017clickbait}. They use a logistic regression classifier for detecting the stance of headlines concerning the body of their articles. Various word level features including punctuation marks and part of speech tags were used to train an SVM classifier for predicting satire and real news articles with 82\% accuracy \cite{rubin2016fake}. 

Scholars have also been able to detect misinformation based on features extracted from their visual information. A few studies so far have attempted to detect misinformation based on image information. Gupta et al. used a dataset of ``fake'' images distributed during hurricane sandy \cite{gupta2013faking}. The authors used only non-visual features such as URLs, propagation patterns, user features, as well as tweet features. Using these features, they were able to achieve high accuracy in predicting fake images from real images, although the authors discuss that this high accuracy might be due to the similarity of many images. Another notable study uses visual attributes from tweet images to classify real and fake news images.

Another notable example is research by Jin et al., focuses on features extracted from images \cite{jin2017novel}. In their study, the authors use images in the context of news events. They define news events as mentions of certain keywords in a specific time span. They then extracts a set of features from images including visual clarity (distribution difference between two image sets), Visual Coherence (how coherent images in specific news event are), Visual Similarity (pairwise similarity distribution histogram in an event), visual diversity (visual difference in the image set of a target news event), and Visual Clustering (number of clusters in a news event). They use four classification models: SVM, Logistic Regression, KStar, and Random forest. The authors were able to achieve an accuracy of 83.6 percent using the Random Forest algorithm which hints towards the promise of using image features from misinformation. However, in non-computer science literature, it has been shown that many other more complex contextual features might be good signals for deception and fake news including facial features and emotions \cite{messaris2001role}.

Another approach to analyzing the content of misinformation is through patterns of diffusion and propagation. These approaches are often conducted through temporal analysis of news sharing or through building different types of networks \cite{shu2017fake}. Liang Wu and Huan Liu developed a message characterization method that focuses purely on content propagation \cite{wu2018tracing}. They propose an LSTM-RNN model that purely uses social proximity and community structures as features to characterize fake news. In another study, unsupervised topic models were used to construct a network of stance-based network and mining conflicting viewpoints; while using an iterative method, the authors would score characterize news with high levels of conflicting topics\cite{jin2016news}. Also without looking at the verbal content of news, Ma et al. argue that focusing on temporal patterns and time series of news spread can be an essential feature in detecting rumors \cite{ma2015detect}. Their method, called Dynamic-Series-Time Structure explores the variation of various social context features over time and can be used to detect suspicious news from credible news.

These examples, are among the methods of using news content or relationships between content to categorize specific pieces of news as misinformation or not. These methods include supervised models, network models, network and temporal models, and they use both text and image features to achieve the goal of categorizing news content. As mentioned in the previous sections, the source and consumers of misinformation are also essential elements in the process of misinformation. In the next section, we study computational approaches in categorizing and understanding sources of misinformation.

\subsection{Measuring and categorizing trustworthiness and veracity of sources\label{source}}

As emphasized by Lazer et al. focusing on sources of misinformation can be more useful in combating misinformation because often repeating misinformation even in the context of fact-checking might prove to be damaging \cite{lazer2018science,pennycook2018prior}. Even though the majority of computational approaches specifically focus on content, various scholars have put their primary focus on understanding and categorizing sources of misinformation. 

One of the most significant red flags in regards to sources of misinformation is whether the sources use automated methods to propagate news. Social bots amplify the reach of misinformation and are built to exploit consumers' cognitive and social biases \cite{lazer2017combating}. These bots are built with different intents. They are often benign and have been proved to be helpful in certain situations \cite{ferrara2016rise}. However, more often than not, bots distribute information without verification and can result in circulating false accusations \cite{gupta20131}. In some cases, bots have been shown to focus on specific topics supporting specific political parties or candidates \cite{ratkiewicz2011detecting}. Moreover, bots can behave similarly to humans by liking and sharing news as well as communicating with humans \cite{hwang2012socialbots}. These behaviors and features have been used to develop many different computational methods to detect and battle social bots.

Ferrera et al. propose a taxonomy of social bot detection systems that include graph-based, crowdsourced, and feature-based bot detection methods  \cite{ferrara2016rise}. Graph-based methods often take advantage of the fact that malicious bot accounts are highly connected to other malicious accounts and have used community detection algorithms to detect clusters of social bots. However, in networks with well-formed clusters, these community detection methods perform poorly \cite{viswanath2011analysis}. Alvisi et al. take note of these shortcomings and offer a community detection that takes a local approach rather than a global community detection algorithm that is more resilient to real-world social bot clusters \cite{alvisi2013sok}. Crowd-sourced methods assume a human's ability to differentiate social bots and legitimate accounts \cite{ferrara2016rise}. Wang et al. conduct a study using both experts and Mechanical Turk and find that while Mechanical Turk workers vary in their efficiency, experts achieve "near-optimal" accuracy in detecting bots \cite{wang2012social}. This is the reason why many large companies hire groups of experts to take charge of detecting social bots. However, using crowd-sourced information is not always cost-effective. The Feature-based category focuses on feature engineering from bots' behaviors and content and uses Machine Learning to predict Social Bots. ``Bot or Not?'' is a well-known example that utilizes such methods \cite{davis2016botornot}. The system uses a combination of linguistic and sentiment features with a Random Forest model to score the likelihood of a Twitter account being a bot or not.  

Other scholars have developed computational methods to go beyond the concept of social bots. Less concerned with misinformation, in an attempt to help journalism experts detect and assess sources of misinformation at the time of breaking news, Diakopoulos et al. create two classifiers. The first classifier categorizes sources into organizations, journalists/blogger and ordinary people \cite{diakopoulos2012finding}. A second classifier uses a dictionary-based approach to identify eyewitnesses and achieves low false-positive rates (89\% precision) but a high false negative rate (32\% recall). The authors provide other cues to help users with the task of source classification including named entities, URL categorization, and spatial information \cite{diakopoulos2012finding}. Castillo et al. combine features from the content of tweets with source level features such as registration age. Statuses count, the number of followers and friends, their verified status, and the existence of description and URL in their user profile to classify news accounts into credible or not \cite{castillo2011information}. Other source based features such as the location of users and also the client used to produce news are also shown to be useful in predicting the credibility of sources \cite{yang2012automatic}. 

Volkova et al. created a neural network classifier that utilizes various linguistic features including Biased Language, Subjectivity, and Moral Foundations, along with signals from the news sources' social activities on Twitter to predict Suspicious Vs. Real news as well as multiple misinformation categories such as clickbaits, Hoaxes, and propaganda \cite{volkova2017separating}. They developed a Convolutional Neural Network model that achieves high levels of accuracy at the binary classification of misinformation sources (real vs. fake). In their model, including mention/retweet interactions between sources greatly improved the performance of the results. The authors also found differences in language cues between different types of accounts. For example, ``verified'' news sources contain significantly fewer markers for bias language, as well as harm, loyalty, and authority moral cues. The authors also find that users retweeting misinformation sources, send high volumes of tweets of shorter periods of time. Inspired by the findings of Volkova et al. \cite{volkova2017separating}, Karduni et al. developed a model using a random forest classifier to separate misinformation from real news and found features such as fear, anger, and negativity to be highly correlated with misinformation account, while features such as fairness and loyalty were highly correlated with real news accounts \cite{karduni2018can}. 

Characterizing sources of misinformation has been the primary goal of various computational methods. Network-based, crowd-sourced based and feature based methods have been used to detect malicious social bots. Others focused on building classifiers using source related features that can categorize other sources of misinformation based on their behavior and usage of content. Developing robust methods to detect and characterize the source and content of misinformation is extremely valuable. However, the question of whether these methods can be useful to reduce the harm of misinformation remains unanswered. Some user-facing tools have been created that aim to assist general users or experts in detecting and rebuking misinformation. These systems were evaluated based on different criteria. In the next section, I move to the third element in the triad of misinformation. By focusing on Consumers, I offer an overview of the existing user-facing systems and their evaluation attempts.

\subsection{Human-Misinformation interaction: Interactive tools, Visual Analytic Systems, Cognitive bias mitigation efforts\label{Consumer}}

The vast majority of computational efforts on battling misinformation has been in detecting misinformation from content and source perspective. However, arguably, the ultimate goal of misinformation mitigation and detection efforts would be to help reduce the effect of the information on users. It has been suggested that providing automated or authoritative fact-checking results along with misinformation might prove counterproductive \cite{lazer2018science, lazer2017combating}. There have been numerous studies in the fields of psychology, cognitive science, and social sciences that highlight the complexities of the relationship between humans and misinformation \cite{pennycook2018prior,pennycook2018cognitive,pennycook2018falls,allcott2017social,arsenault2006conquering}. However, the efforts in the fields of computation have mostly neglected the human aspects dealing with information. In this section, I offer a highlight of some of the notable efforts in human-computer-interaction (HCI). Moreover, I also introduce notable interactive interfaces built to combat misinformation. 

Some studies have examined how users ability to detect misinformation while using computational tools. Flintham et al. conducted a survey and set of interviews to understand consumer behavior and attitude, as well as their strategies to detect misinformation on social media. Within their study, they ask users to find "fake news" on Facebook while thinking aloud. Using a thematic analysis of qualitative data, they found that users had different approaches towards sources of misinformation, some giving complete primacy to the authenticity of the source, while others decided to disregard the source’s reputation. The authors also found that authors interest in a specific "kind" of news played an important role in their interest in separating facts from fake news. Finally, they found that their participants were reluctant to trust a tool that would allow them to determine the veracity of misinformation on social media \cite{flintham2018falling}. Pourghomi et al. study the interaction techniques utilizes by Facebook to help users fact-check news posts and compare those methods with another proposed method called "right click authenticate".  The method presents facts related and editorial pieces formatted similar to Wikipedia and suggests that it might be more useful to engage in authenticating news themselves rather than relying on third-party fact-checkers \cite{pourghomi2017stop}. In another notable study, Kasra et al. conduct a focus group study on users to understand how well they can detect and rebuke doctored and faked images. Their findings suggest that users do not perform well at identifying fake online images. They also found that the users’ main strategy to rebuke images was to refer to non-visual attributes such as the source and the accompanying description. Moreover, they found that users failed to identify cues in images when specifically asked for \cite{kasra2018seeing}. 

To build a visual system to help journalists asses sources of news, Diakopoulos et al. develop an interface specifically built for journalism experts. The interface offers a variety of information regarding sources including the number of friends and followers, the location they mostly tweet from, Whether the account is an eyewitness to specific news, and named entities extracted from the content of the tweets. They conducted a series of expert interviews and found intriguing results concerning their needs. Even though the system was not built explicitly for sources of misinformation, the interviewees expressed their interest in features to detect misinformation \cite{diakopoulos2012finding}. Narwal et al. develop an automated assistant called UnbiassedCrowd that aims to help users understand biases in visual information on Twitter. The system collects images, clusters them using by extracting fisher vectors and K-Means clustering. The system then allows users to highlight biases and share to others using automated bots or manually. The authors conducted a study with experts and the general public. One of the interesting findings was the need for providing context to clustered images. Moreover, their general public study found two groups, one who actively propagated the information about biased images and a group that took a defensive stance which highlights users different strategies towards content verification \cite{narwal2017automated}.

Several systems and interfaces have been developed with varying amounts of interactivity. Gupta et al. develop TweetCred that provides real-time credibility assessment of content on Twitter. The system uses an SVM classifier on produces a visual score between 1 and 7 for the credibility of each news content on Twitter. The system does not offer other interactive tools for users to deal with misinformation\cite{gupta2014tweetcred}. Emergent is a dataset of rumors and a real-time rumor tracking website. The system mostly serves as a fact-checking source that offers fact-checking by simply tagging claims as True, False, or unverified and offering users extra information such as the originating source, number of shares, and topic tags \cite{Emergent90:online}. 

Twittertrails is an interactive web-based system that affords users to view fact-checking information on specific claims. The system allows for searching based on keywords, category, and levels of spread and skepticism. Linearly, each claim is tagged with two radial encoding, one for the spread and one for skepticism. Selecting each rumor opens a new page that provides different interactive visualizations and discussions on propagation, temporal nature, level of visibility, topics related to the rumor, and images used in tweets related to the rumor. The system does not go beyond fact-checking to test usability and relationship to how consumers would use the system\cite{TwitterT44:online}. RumorLens is one of the first systems that combine computational tools such as keyword-based classifiers for rumors with interactive visualizations to help humans in the task of rumor detection. The system includes a network diagram for highlighting the propagation of rumors, a snakey diagram paired with a timeline of the lifetime of the rumor. The system allows users to explore the progression of a rumor detected from twitter using a combination of different visualizations \cite{resnick2014rumorlens}. Hoaxy is a search engine and a dashboard that combines various scores on misinformation accounts, as well as interactive visualizations of the timeline of propagation and a network visualization to allow users to explore misinformation on social media. Hoaxy offers information on Accounts that share misinformation as well as information about the content of misinformation from these accounts \cite{shao2016hoaxy}. Finally, RumorFlow is a Visual Analytics tool for understanding and analyzing how rumors are disseminated and discussed by users of social media. It uses Reddit as a source for data. It uses semantic similarity, sentiment analysis, and Wikipedia Entity Linking as methods of extracting extra information form the content. The visual analytics system includes a theme river visualization to highlight the development of rumors, a word cloud, and topic cloud, as well as a snakey diagram showing relationships between topics related to rumors \cite{dang2016rumour}. 

Most of these systems and studies share a common goal: they provide preliminary studies on how consumers approach misinformation and offer computational tools in the form of interfaces that offer a combination of automated scores, basic visualizations, and contextual information. However, the majority of the efforts do not include efforts to understand the usefulness of the system. Moreover, none of these efforts go in-depth into the issues that cause consumers to be affected by misinformation. Overall, there seems to be a real gap in computer science literature that utilize computational efforts to help the complex decision-making process of consumers. Future efforts, require careful studies on how computational methods and tools can be used in a real-world context to help consumers make informed decisions about misinformation online.

In the next section, I offer some discussions on the overall landscape of misinformation research in computer science and information technology and conclude with remarks on some potential interdisciplinary research paths for combating misinformation. 

\section{discussion and conclusion\label{discussion}}

Studying computational methods of battling misinformation through a systematic framework with three main elements of Source, Content, and Consumer allows us to categorize these approaches into three categories of work. The first category, which comprises majority of computational effort, is misinformation detection and verification. These works generally focus purely on determining the veracity of news content. The work in this category generally focus on a combination of tasks: To Automatically fact-check claims and provide scores, context, alternative claims, etc; and to detect and classify verbal and visual misinformation of various kinds ( deception, clickbaits, rumors, propaganda, etc ). Automatic Fact-checking which has proved to be a very difficult task \cite{graves2018understanding} can be done though crowdsourcing, checking with external sources, or a combination of classifiers and knowledge-bases \cite{graves2018understanding}. On the other hand, detection approaches have used a variety of methods  to distinguish misinformation. These methods include using language and styles and features as a signal, using visual features, and through temporal and propagation patterns of misinformation.

Another group of studies focus on determining and verifying sources of misinformation. As one of the biggest modern challenges of misinformation is utilization of social bots, in this category of work, a lot of effort has been put on developing methods to automatically classify sources as bot or human accounts. Bot detection is approached using graph-based methods, crowd-sourced methods, and by creating classifiers that use sources' behavior and aggregate content as features. A group of scholars have also developed methods to score and classify the trustworthiness of news sources. These works utilize aggregate signals from the content of different sources including language and writing styles, features extracted from social network behavior of sources, as well as other metadata such as location and registration age.

The third category which arguably can be the most important one, is consumer-facing systems and studies. These works aim to provide tools to users to understand, detect, and mitigate the effects of misinformation. A number of interactive systems have been developed that mostly allow users to fact-check misinformation as well as to visualize the behavior of sources. Empirical studies on the usefulness and efficacy of these systems and methods are extremely sparse. Moreover, the existing studies do not deal with findings from other disciplines that are known to be extremely important in the process of battling misinformation. These findings include confirmation biases, prior exposure, social and peer pressure, echo-chambers and filter bubbles. 

Even though the potential of using computational methods to battle misinformation has been discussed \cite{lazer2017combating,lazer2018science}, there still hasn't been studies that provide insights on how computational tools and visual systems can be used to moderate the effects of multiple cognitive biases and social pressures. Even though there has been great advancements in computational detection of misinformation content and sources, in order to truly battle misinformation through computational tools, one of the biggest next steps should be to develop methods and tools that are designed to deal with the complex psychological and social process of how humans consume misinformation and why they are effected by it. This, in fact, requires a collaborative interdisciplinary attempt bringing together experts from psychology, cognitive sciences, education, social and political sciences, and computer science.

%
\bibliographystyle{ACM-Reference-Format}
\bibliography{sample-base}


\begin{thebibliography}{121}


\ifx \showCODEN    \undefined \def \showCODEN     #1{\unskip}     \fi
\ifx \showDOI      \undefined \def \showDOI       #1{#1}\fi
\ifx \showISBNx    \undefined \def \showISBNx     #1{\unskip}     \fi
\ifx \showISBNxiii \undefined \def \showISBNxiii  #1{\unskip}     \fi
\ifx \showISSN     \undefined \def \showISSN      #1{\unskip}     \fi
\ifx \showLCCN     \undefined \def \showLCCN      #1{\unskip}     \fi
\ifx \shownote     \undefined \def \shownote      #1{#1}          \fi
\ifx \showarticletitle \undefined \def \showarticletitle #1{#1}   \fi
\ifx \showURL      \undefined \def \showURL       {\relax}        \fi
\providecommand\bibfield[2]{#2}
\providecommand\bibinfo[2]{#2}
\providecommand\natexlab[1]{#1}
\providecommand\showeprint[2][]{arXiv:#2}

\bibitem[\protect\citeauthoryear{??}{Fac}{[n. d.]a}]%
        {Facebook81:online}
 \bibinfo{year}{[n. d.]}\natexlab{a}.
\newblock \bibinfo{title}{| Facebook Media and Publisher Help Center}.
\newblock
  \bibinfo{howpublished}{\url{https://www.facebook.com/help/publisher/182222309230722}}.
\newblock
\newblock
\shownote{(Accessed on 01/22/2019).}


\bibitem[\protect\citeauthoryear{??}{Eme}{[n. d.]}]%
        {Emergent90:online}
 \bibinfo{year}{[n. d.]}\natexlab{}.
\newblock \bibinfo{title}{Emergent}.
\newblock \bibinfo{howpublished}{\url{http://www.emergent.info/}}.
\newblock
\newblock
\shownote{(Accessed on 02/07/2019).}


\bibitem[\protect\citeauthoryear{??}{Fac}{[n. d.]b}]%
        {FactChec34:online}
 \bibinfo{year}{[n. d.]}\natexlab{b}.
\newblock \bibinfo{title}{Fact- Checking - Duke Reporters' Lab}.
\newblock
  \bibinfo{howpublished}{\url{https://reporterslab.org/fact-checking/}}.
\newblock
\newblock
\shownote{(Accessed on 01/22/2019).}


\bibitem[\protect\citeauthoryear{??}{sno}{[n. d.]}]%
        {snopes.com_2018}
 \bibinfo{year}{[n. d.]}\natexlab{}.
\newblock \bibinfo{booktitle}{\emph{Fact-checking U.S. politics}}.
\newblock
\urldef\tempurl%
\url{http://www.Snopes.com/}
\showURL{%
\tempurl}


\bibitem[\protect\citeauthoryear{??}{Fac}{[n. d.]c}]%
        {Factchec51:online}
 \bibinfo{year}{[n. d.]}\natexlab{c}.
\newblock \bibinfo{title}{Fact-checking U.S. politics | PolitiFact}.
\newblock \bibinfo{howpublished}{\url{https://www.politifact.com/}}.
\newblock
\newblock
\shownote{(Accessed on 01/22/2019).}


\bibitem[\protect\citeauthoryear{??}{Fac}{[n. d.]d}]%
        {FactChec59:online}
 \bibinfo{year}{[n. d.]}\natexlab{d}.
\newblock \bibinfo{title}{FactCheck.org - A Project of The Annenberg Public
  Policy Center}.
\newblock \bibinfo{howpublished}{\url{https://www.factcheck.org/}}.
\newblock
\newblock
\shownote{(Accessed on 01/22/2019).}


\bibitem[\protect\citeauthoryear{??}{Fis}{[n. d.]}]%
        {Fiskkitc60:online}
 \bibinfo{year}{[n. d.]}\natexlab{}.
\newblock \bibinfo{title}{Fiskkit.com Discuss news that matters and find out
  what's true.}
\newblock \bibinfo{howpublished}{\url{http://fiskkit.com/}}.
\newblock
\newblock
\shownote{(Accessed on 01/22/2019).}


\bibitem[\protect\citeauthoryear{??}{Goo}{[n. d.]}]%
        {GoogleFa66:online}
 \bibinfo{year}{[n. d.]}\natexlab{}.
\newblock \bibinfo{title}{Google, Facebook and Twitter Agree to Fight Fake News
  in the EU - Bloomberg}.
\newblock
  \bibinfo{howpublished}{\url{https://www.bloomberg.com/news/articles/2018-09-25/google-facebook-and-twitter-agree-to-fight-fake-news-in-eu}}.
\newblock
\newblock
\shownote{(Accessed on 01/22/2019).}


\bibitem[\protect\citeauthoryear{??}{Lat}{[n. d.]}]%
        {LatestEm20:online}
 \bibinfo{year}{[n. d.]}\natexlab{}.
\newblock \bibinfo{title}{Latest Email Hoaxes - Current Internet Scams -
  Hoax-Slayer}.
\newblock \bibinfo{howpublished}{\url{https://hoax-slayer.com/}}.
\newblock
\newblock
\shownote{(Accessed on 01/22/2019).}


\bibitem[\protect\citeauthoryear{??}{Thi}{[n. d.]}]%
        {Thisisho46:online}
 \bibinfo{year}{[n. d.]}\natexlab{}.
\newblock \bibinfo{title}{This is how Facebook's news feed fact-checking will
  work in the UK | WIRED UK}.
\newblock
  \bibinfo{howpublished}{\url{https://www.wired.co.uk/article/full-fact-facebook-fact-checking}}.
\newblock
\newblock
\shownote{(Accessed on 01/22/2019).}


\bibitem[\protect\citeauthoryear{??}{Twi}{[n. d.]}]%
        {TwitterT44:online}
 \bibinfo{year}{[n. d.]}\natexlab{}.
\newblock \bibinfo{title}{Twitter Trails: Tool for monitoring the propagation
  of rumors}.
\newblock \bibinfo{howpublished}{\url{http://twittertrails.com/}}.
\newblock
\newblock
\shownote{(Accessed on 02/07/2019).}


\bibitem[\protect\citeauthoryear{??}{Why}{[n. d.]}]%
        {WhytheMa1:online}
 \bibinfo{year}{[n. d.]}\natexlab{}.
\newblock \bibinfo{title}{Why the Macron Hacking Attack Landed With a Thud in
  France - The New York Times}.
\newblock
  \bibinfo{howpublished}{\url{https://www.nytimes.com/2017/05/08/world/europe/macron-hacking-attack-france.html}}.
\newblock
\newblock
\shownote{(Accessed on 12/02/2018).}


\bibitem[\protect\citeauthoryear{Abraham and Appiah}{Abraham and
  Appiah}{2006}]%
        {abraham2006framing}
\bibfield{author}{\bibinfo{person}{Linus Abraham} {and} \bibinfo{person}{Osei
  Appiah}.} \bibinfo{year}{2006}\natexlab{}.
\newblock \showarticletitle{Framing news stories: The role of visual imagery in
  priming racial stereotypes}.
\newblock \bibinfo{journal}{\emph{The Howard Journal of Communications}}
  \bibinfo{volume}{17}, \bibinfo{number}{3} (\bibinfo{year}{2006}),
  \bibinfo{pages}{183--203}.
\newblock


\bibitem[\protect\citeauthoryear{Adams}{Adams}{1986}]%
        {adams1986whose}
\bibfield{author}{\bibinfo{person}{William~C Adams}.}
  \bibinfo{year}{1986}\natexlab{}.
\newblock \showarticletitle{Whose lives count? TV coverage of natural
  disasters}.
\newblock \bibinfo{journal}{\emph{Journal of Communication}}
  \bibinfo{volume}{36}, \bibinfo{number}{2} (\bibinfo{year}{1986}),
  \bibinfo{pages}{113--122}.
\newblock


\bibitem[\protect\citeauthoryear{Allcott and Gentzkow}{Allcott and
  Gentzkow}{2017}]%
        {allcott2017social}
\bibfield{author}{\bibinfo{person}{Hunt Allcott} {and} \bibinfo{person}{Matthew
  Gentzkow}.} \bibinfo{year}{2017}\natexlab{}.
\newblock \showarticletitle{Social media and fake news in the 2016 election}.
\newblock \bibinfo{journal}{\emph{Journal of Economic Perspectives}}
  \bibinfo{volume}{31}, \bibinfo{number}{2} (\bibinfo{year}{2017}),
  \bibinfo{pages}{211--36}.
\newblock


\bibitem[\protect\citeauthoryear{Alvisi, Clement, Epasto, Lattanzi, and
  Panconesi}{Alvisi et~al\mbox{.}}{2013}]%
        {alvisi2013sok}
\bibfield{author}{\bibinfo{person}{Lorenzo Alvisi}, \bibinfo{person}{Allen
  Clement}, \bibinfo{person}{Alessandro Epasto}, \bibinfo{person}{Silvio
  Lattanzi}, {and} \bibinfo{person}{Alessandro Panconesi}.}
  \bibinfo{year}{2013}\natexlab{}.
\newblock \showarticletitle{Sok: The evolution of sybil defense via social
  networks}. In \bibinfo{booktitle}{\emph{Security and Privacy (SP), 2013 IEEE
  Symposium on}}. IEEE, \bibinfo{pages}{382--396}.
\newblock


\bibitem[\protect\citeauthoryear{Arsenault and Castells}{Arsenault and
  Castells}{2006}]%
        {arsenault2006conquering}
\bibfield{author}{\bibinfo{person}{Amelia Arsenault} {and}
  \bibinfo{person}{Manuel Castells}.} \bibinfo{year}{2006}\natexlab{}.
\newblock \showarticletitle{Conquering the minds, conquering Iraq: The social
  production of misinformation in the United States--a case study}.
\newblock \bibinfo{journal}{\emph{Information, Communication \& Society}}
  \bibinfo{volume}{9}, \bibinfo{number}{3} (\bibinfo{year}{2006}),
  \bibinfo{pages}{284--307}.
\newblock


\bibitem[\protect\citeauthoryear{Bakir and McStay}{Bakir and McStay}{2018}]%
        {bakir2018fake}
\bibfield{author}{\bibinfo{person}{Vian Bakir} {and} \bibinfo{person}{Andrew
  McStay}.} \bibinfo{year}{2018}\natexlab{}.
\newblock \showarticletitle{Fake news and the economy of emotions: Problems,
  causes, solutions}.
\newblock \bibinfo{journal}{\emph{Digital Journalism}} \bibinfo{volume}{6},
  \bibinfo{number}{2} (\bibinfo{year}{2018}), \bibinfo{pages}{154--175}.
\newblock


\bibitem[\protect\citeauthoryear{Bakshy, Messing, and Adamic}{Bakshy
  et~al\mbox{.}}{2015}]%
        {bakshy2015exposure}
\bibfield{author}{\bibinfo{person}{Eytan Bakshy}, \bibinfo{person}{Solomon
  Messing}, {and} \bibinfo{person}{Lada~A Adamic}.}
  \bibinfo{year}{2015}\natexlab{}.
\newblock \showarticletitle{Exposure to ideologically diverse news and opinion
  on Facebook}.
\newblock \bibinfo{journal}{\emph{Science}} \bibinfo{volume}{348},
  \bibinfo{number}{6239} (\bibinfo{year}{2015}), \bibinfo{pages}{1130--1132}.
\newblock


\bibitem[\protect\citeauthoryear{Barber{\'a}, Jost, Nagler, Tucker, and
  Bonneau}{Barber{\'a} et~al\mbox{.}}{2015}]%
        {barbera2015tweeting}
\bibfield{author}{\bibinfo{person}{Pablo Barber{\'a}}, \bibinfo{person}{John~T
  Jost}, \bibinfo{person}{Jonathan Nagler}, \bibinfo{person}{Joshua~A Tucker},
  {and} \bibinfo{person}{Richard Bonneau}.} \bibinfo{year}{2015}\natexlab{}.
\newblock \showarticletitle{Tweeting from left to right: Is online political
  communication more than an echo chamber?}
\newblock \bibinfo{journal}{\emph{Psychological science}} \bibinfo{volume}{26},
  \bibinfo{number}{10} (\bibinfo{year}{2015}), \bibinfo{pages}{1531--1542}.
\newblock


\bibitem[\protect\citeauthoryear{Baym and Jones}{Baym and Jones}{2013}]%
        {baym2013news}
\bibfield{author}{\bibinfo{person}{Geoffrey Baym} {and}
  \bibinfo{person}{Jeffrey~P Jones}.} \bibinfo{year}{2013}\natexlab{}.
\newblock \bibinfo{booktitle}{\emph{News parody and political satire across the
  globe}}.
\newblock \bibinfo{publisher}{Routledge}.
\newblock


\bibitem[\protect\citeauthoryear{Benkler, Faris, and Roberts}{Benkler
  et~al\mbox{.}}{2018}]%
        {benkler2018network}
\bibfield{author}{\bibinfo{person}{Yochai Benkler}, \bibinfo{person}{Robert
  Faris}, {and} \bibinfo{person}{Hal Roberts}.}
  \bibinfo{year}{2018}\natexlab{}.
\newblock \bibinfo{booktitle}{\emph{Network Propaganda: Manipulation,
  Disinformation, and Radicalization in American Politics}}.
\newblock \bibinfo{publisher}{Oxford University Press}.
\newblock


\bibitem[\protect\citeauthoryear{Bessi and Ferrara}{Bessi and Ferrara}{2016}]%
        {bessi2016social}
\bibfield{author}{\bibinfo{person}{Alessandro Bessi} {and}
  \bibinfo{person}{Emilio Ferrara}.} \bibinfo{year}{2016}\natexlab{}.
\newblock \showarticletitle{Social bots distort the 2016 US Presidential
  election online discussion}.
\newblock  (\bibinfo{year}{2016}).
\newblock


\bibitem[\protect\citeauthoryear{Bourgonje, Schneider, and Rehm}{Bourgonje
  et~al\mbox{.}}{2017}]%
        {bourgonje2017clickbait}
\bibfield{author}{\bibinfo{person}{Peter Bourgonje},
  \bibinfo{person}{Julian~Moreno Schneider}, {and} \bibinfo{person}{Georg
  Rehm}.} \bibinfo{year}{2017}\natexlab{}.
\newblock \showarticletitle{From clickbait to fake news detection: an approach
  based on detecting the stance of headlines to articles}. In
  \bibinfo{booktitle}{\emph{Proceedings of the 2017 EMNLP Workshop: Natural
  Language Processing meets Journalism}}. \bibinfo{pages}{84--89}.
\newblock


\bibitem[\protect\citeauthoryear{Carlson}{Carlson}{2009}]%
        {carlson2009reality}
\bibfield{author}{\bibinfo{person}{Matt Carlson}.}
  \bibinfo{year}{2009}\natexlab{}.
\newblock \showarticletitle{THE REALITY OF A FAKE IMAGE News norms,
  photojournalistic craft, and Brian Walski's fabricated photograph}.
\newblock \bibinfo{journal}{\emph{Journalism Practice}} \bibinfo{volume}{3},
  \bibinfo{number}{2} (\bibinfo{year}{2009}), \bibinfo{pages}{125--139}.
\newblock


\bibitem[\protect\citeauthoryear{Castillo, Mendoza, and Poblete}{Castillo
  et~al\mbox{.}}{2011}]%
        {castillo2011information}
\bibfield{author}{\bibinfo{person}{Carlos Castillo}, \bibinfo{person}{Marcelo
  Mendoza}, {and} \bibinfo{person}{Barbara Poblete}.}
  \bibinfo{year}{2011}\natexlab{}.
\newblock \showarticletitle{Information credibility on twitter}. In
  \bibinfo{booktitle}{\emph{Proceedings of the 20th international conference on
  World wide web}}. ACM, \bibinfo{pages}{675--684}.
\newblock


\bibitem[\protect\citeauthoryear{Chen, Sin, Theng, and Lee}{Chen
  et~al\mbox{.}}{2015}]%
        {chen2015students}
\bibfield{author}{\bibinfo{person}{Xinran Chen},
  \bibinfo{person}{Sei-Ching~Joanna Sin}, \bibinfo{person}{Yin-Leng Theng},
  {and} \bibinfo{person}{Chei~Sian Lee}.} \bibinfo{year}{2015}\natexlab{}.
\newblock \showarticletitle{Why students share misinformation on social media:
  Motivation, gender, and study-level differences}.
\newblock \bibinfo{journal}{\emph{The Journal of Academic Librarianship}}
  \bibinfo{volume}{41}, \bibinfo{number}{5} (\bibinfo{year}{2015}),
  \bibinfo{pages}{583--592}.
\newblock


\bibitem[\protect\citeauthoryear{Coleman}{Coleman}{2010}]%
        {coleman2010framing}
\bibfield{author}{\bibinfo{person}{Renita Coleman}.}
  \bibinfo{year}{2010}\natexlab{}.
\newblock \showarticletitle{Framing the pictures in our heads}.
\newblock \bibinfo{journal}{\emph{Doing news framing analysis: Empirical and
  theoretical perspectives}} (\bibinfo{year}{2010}), \bibinfo{pages}{233--261}.
\newblock


\bibitem[\protect\citeauthoryear{Dang, Moh'd, Milios, and Minghim}{Dang
  et~al\mbox{.}}{2016}]%
        {dang2016rumour}
\bibfield{author}{\bibinfo{person}{Anh Dang}, \bibinfo{person}{Abidalrahman
  Moh'd}, \bibinfo{person}{Evangelos Milios}, {and} \bibinfo{person}{Rosane
  Minghim}.} \bibinfo{year}{2016}\natexlab{}.
\newblock \showarticletitle{What is in a rumour: Combined visual analysis of
  rumour flow and user activity}. In \bibinfo{booktitle}{\emph{Proceedings of
  the 33rd Computer Graphics International}}. ACM, \bibinfo{pages}{17--20}.
\newblock


\bibitem[\protect\citeauthoryear{Davis, Varol, Ferrara, Flammini, and
  Menczer}{Davis et~al\mbox{.}}{2016}]%
        {davis2016botornot}
\bibfield{author}{\bibinfo{person}{Clayton~Allen Davis}, \bibinfo{person}{Onur
  Varol}, \bibinfo{person}{Emilio Ferrara}, \bibinfo{person}{Alessandro
  Flammini}, {and} \bibinfo{person}{Filippo Menczer}.}
  \bibinfo{year}{2016}\natexlab{}.
\newblock \showarticletitle{Botornot: A system to evaluate social bots}. In
  \bibinfo{booktitle}{\emph{Proceedings of the 25th International Conference
  Companion on World Wide Web}}. International World Wide Web Conferences
  Steering Committee, \bibinfo{pages}{273--274}.
\newblock


\bibitem[\protect\citeauthoryear{Day and Thompson}{Day and Thompson}{2012}]%
        {day2012live}
\bibfield{author}{\bibinfo{person}{Amber Day} {and} \bibinfo{person}{Ethan
  Thompson}.} \bibinfo{year}{2012}\natexlab{}.
\newblock \showarticletitle{Live from New York, it's the fake news! Saturday
  Night Live and the (non) politics of parody}.
\newblock \bibinfo{journal}{\emph{Popular Communication}} \bibinfo{volume}{10},
  \bibinfo{number}{1-2} (\bibinfo{year}{2012}), \bibinfo{pages}{170--182}.
\newblock


\bibitem[\protect\citeauthoryear{Diakopoulos, De~Choudhury, and
  Naaman}{Diakopoulos et~al\mbox{.}}{2012}]%
        {diakopoulos2012finding}
\bibfield{author}{\bibinfo{person}{Nicholas Diakopoulos},
  \bibinfo{person}{Munmun De~Choudhury}, {and} \bibinfo{person}{Mor Naaman}.}
  \bibinfo{year}{2012}\natexlab{}.
\newblock \showarticletitle{Finding and assessing social media information
  sources in the context of journalism}. In
  \bibinfo{booktitle}{\emph{Proceedings of the SIGCHI Conference on Human
  Factors in Computing Systems}}. ACM, \bibinfo{pages}{2451--2460}.
\newblock


\bibitem[\protect\citeauthoryear{Dong, Gabrilovich, Heitz, Horn, Lao, Murphy,
  Strohmann, Sun, and Zhang}{Dong et~al\mbox{.}}{2014}]%
        {dong2014knowledge}
\bibfield{author}{\bibinfo{person}{Xin Dong}, \bibinfo{person}{Evgeniy
  Gabrilovich}, \bibinfo{person}{Geremy Heitz}, \bibinfo{person}{Wilko Horn},
  \bibinfo{person}{Ni Lao}, \bibinfo{person}{Kevin Murphy},
  \bibinfo{person}{Thomas Strohmann}, \bibinfo{person}{Shaohua Sun}, {and}
  \bibinfo{person}{Wei Zhang}.} \bibinfo{year}{2014}\natexlab{}.
\newblock \showarticletitle{Knowledge vault: A web-scale approach to
  probabilistic knowledge fusion}. In \bibinfo{booktitle}{\emph{Proceedings of
  the 20th ACM SIGKDD international conference on Knowledge discovery and data
  mining}}. ACM, \bibinfo{pages}{601--610}.
\newblock


\bibitem[\protect\citeauthoryear{Eckles and Bakshy}{Eckles and Bakshy}{2017}]%
        {eckles2017bias}
\bibfield{author}{\bibinfo{person}{Dean Eckles} {and} \bibinfo{person}{Eytan
  Bakshy}.} \bibinfo{year}{2017}\natexlab{}.
\newblock \showarticletitle{Bias and high-dimensional adjustment in
  observational studies of peer effects}.
\newblock \bibinfo{journal}{\emph{arXiv preprint arXiv:1706.04692}}
  (\bibinfo{year}{2017}).
\newblock


\bibitem[\protect\citeauthoryear{Entman}{Entman}{2010}]%
        {entman2010media}
\bibfield{author}{\bibinfo{person}{Robert~M Entman}.}
  \bibinfo{year}{2010}\natexlab{}.
\newblock \showarticletitle{Media framing biases and political power:
  Explaining slant in news of Campaign 2008}.
\newblock \bibinfo{journal}{\emph{Journalism}} \bibinfo{volume}{11},
  \bibinfo{number}{4} (\bibinfo{year}{2010}), \bibinfo{pages}{389--408}.
\newblock


\bibitem[\protect\citeauthoryear{Feng, Banerjee, and Choi}{Feng
  et~al\mbox{.}}{2012}]%
        {feng2012syntactic}
\bibfield{author}{\bibinfo{person}{Song Feng}, \bibinfo{person}{Ritwik
  Banerjee}, {and} \bibinfo{person}{Yejin Choi}.}
  \bibinfo{year}{2012}\natexlab{}.
\newblock \showarticletitle{Syntactic stylometry for deception detection}. In
  \bibinfo{booktitle}{\emph{Proceedings of the 50th Annual Meeting of the
  Association for Computational Linguistics: Short Papers-Volume 2}}.
  Association for Computational Linguistics, \bibinfo{pages}{171--175}.
\newblock


\bibitem[\protect\citeauthoryear{Ferrara, Varol, Davis, Menczer, and
  Flammini}{Ferrara et~al\mbox{.}}{2016}]%
        {ferrara2016rise}
\bibfield{author}{\bibinfo{person}{Emilio Ferrara}, \bibinfo{person}{Onur
  Varol}, \bibinfo{person}{Clayton Davis}, \bibinfo{person}{Filippo Menczer},
  {and} \bibinfo{person}{Alessandro Flammini}.}
  \bibinfo{year}{2016}\natexlab{}.
\newblock \showarticletitle{The rise of social bots}.
\newblock \bibinfo{journal}{\emph{Commun. ACM}} \bibinfo{volume}{59},
  \bibinfo{number}{7} (\bibinfo{year}{2016}), \bibinfo{pages}{96--104}.
\newblock


\bibitem[\protect\citeauthoryear{Figueira and Oliveira}{Figueira and
  Oliveira}{2017}]%
        {figueira2017current}
\bibfield{author}{\bibinfo{person}{{\'A}lvaro Figueira} {and}
  \bibinfo{person}{Luciana Oliveira}.} \bibinfo{year}{2017}\natexlab{}.
\newblock \showarticletitle{The current state of fake news: challenges and
  opportunities}.
\newblock \bibinfo{journal}{\emph{Procedia Computer Science}}
  \bibinfo{volume}{121} (\bibinfo{year}{2017}), \bibinfo{pages}{817--825}.
\newblock


\bibitem[\protect\citeauthoryear{Flaxman, Goel, and Rao}{Flaxman
  et~al\mbox{.}}{2016}]%
        {flaxman2016filter}
\bibfield{author}{\bibinfo{person}{Seth Flaxman}, \bibinfo{person}{Sharad
  Goel}, {and} \bibinfo{person}{Justin~M Rao}.}
  \bibinfo{year}{2016}\natexlab{}.
\newblock \showarticletitle{Filter bubbles, echo chambers, and online news
  consumption}.
\newblock \bibinfo{journal}{\emph{Public Opinion Quarterly}}
  \bibinfo{volume}{80}, \bibinfo{number}{S1} (\bibinfo{year}{2016}),
  \bibinfo{pages}{298--320}.
\newblock


\bibitem[\protect\citeauthoryear{Flintham, Karner, Bachour, Creswick, Gupta,
  and Moran}{Flintham et~al\mbox{.}}{2018}]%
        {flintham2018falling}
\bibfield{author}{\bibinfo{person}{Martin Flintham}, \bibinfo{person}{Christian
  Karner}, \bibinfo{person}{Khaled Bachour}, \bibinfo{person}{Helen Creswick},
  \bibinfo{person}{Neha Gupta}, {and} \bibinfo{person}{Stuart Moran}.}
  \bibinfo{year}{2018}\natexlab{}.
\newblock \showarticletitle{Falling for fake news: investigating the
  consumption of news via social media}. In
  \bibinfo{booktitle}{\emph{Proceedings of the 2018 CHI Conference on Human
  Factors in Computing Systems}}. ACM, \bibinfo{pages}{376}.
\newblock


\bibitem[\protect\citeauthoryear{Flynn, Nyhan, and Reifler}{Flynn
  et~al\mbox{.}}{2017}]%
        {flynn2017nature}
\bibfield{author}{\bibinfo{person}{DJ Flynn}, \bibinfo{person}{Brendan Nyhan},
  {and} \bibinfo{person}{Jason Reifler}.} \bibinfo{year}{2017}\natexlab{}.
\newblock \showarticletitle{The nature and origins of misperceptions:
  Understanding false and unsupported beliefs about politics}.
\newblock \bibinfo{journal}{\emph{Political Psychology}}  \bibinfo{volume}{38}
  (\bibinfo{year}{2017}), \bibinfo{pages}{127--150}.
\newblock


\bibitem[\protect\citeauthoryear{Frenkel}{Frenkel}{2017}]%
        {frenkel_2017}
\bibfield{author}{\bibinfo{person}{Sheera Frenkel}.}
  \bibinfo{year}{2017}\natexlab{}.
\newblock \showarticletitle{For Russian 'Trolls,' Instagram's Pictures Can
  Spread Wider Than Words}.
\newblock \bibinfo{journal}{\emph{The New York Times}} (\bibinfo{date}{Dec}
  \bibinfo{year}{2017}).
\newblock
\urldef\tempurl%
\url{https://www.nytimes.com/2017/12/17/technology/instagram-russian-trolls.html}
\showURL{%
\tempurl}


\bibitem[\protect\citeauthoryear{Gabbert, Memon, Allan, and Wright}{Gabbert
  et~al\mbox{.}}{2004}]%
        {gabbert2004say}
\bibfield{author}{\bibinfo{person}{Fiona Gabbert}, \bibinfo{person}{Amina
  Memon}, \bibinfo{person}{Kevin Allan}, {and} \bibinfo{person}{Daniel~B
  Wright}.} \bibinfo{year}{2004}\natexlab{}.
\newblock \showarticletitle{Say it to my face: Examining the effects of
  socially encountered misinformation}.
\newblock \bibinfo{journal}{\emph{Legal and Criminological Psychology}}
  \bibinfo{volume}{9}, \bibinfo{number}{2} (\bibinfo{year}{2004}),
  \bibinfo{pages}{215--227}.
\newblock


\bibitem[\protect\citeauthoryear{Grabe and Bucy}{Grabe and Bucy}{2009}]%
        {grabe2009image}
\bibfield{author}{\bibinfo{person}{Maria~Elizabeth Grabe} {and}
  \bibinfo{person}{Erik~Page Bucy}.} \bibinfo{year}{2009}\natexlab{}.
\newblock \bibinfo{booktitle}{\emph{Image bite politics: News and the visual
  framing of elections}}.
\newblock \bibinfo{publisher}{Oxford University Press}.
\newblock


\bibitem[\protect\citeauthoryear{Graves}{Graves}{[n. d.]}]%
        {FACTSHEE77:online}
\bibfield{author}{\bibinfo{person}{Lucas Graves}.} \bibinfo{year}{[n.
  d.]}\natexlab{}.
\newblock \bibinfo{title}{FACTSHEET: Understanding the Promise and Limits of
  Automated Fact-Checking - Reuters Institute Digital News Report}.
\newblock
  \bibinfo{howpublished}{\url{http://www.digitalnewsreport.org/publications/2018/factsheet-understanding-promise-limits-automated-fact-checking/}}.
\newblock
\newblock
\shownote{(Accessed on 01/24/2019).}


\bibitem[\protect\citeauthoryear{Graves}{Graves}{2018}]%
        {graves2018understanding}
\bibfield{author}{\bibinfo{person}{Lucas Graves}.}
  \bibinfo{year}{2018}\natexlab{}.
\newblock \showarticletitle{Understanding the promise and limits of automated
  fact-checking}.
\newblock \bibinfo{journal}{\emph{Factsheet}}  \bibinfo{volume}{2}
  (\bibinfo{year}{2018}), \bibinfo{pages}{2018--02}.
\newblock


\bibitem[\protect\citeauthoryear{Gupta, Kumaraguru, Castillo, and Meier}{Gupta
  et~al\mbox{.}}{2014}]%
        {gupta2014tweetcred}
\bibfield{author}{\bibinfo{person}{Aditi Gupta}, \bibinfo{person}{Ponnurangam
  Kumaraguru}, \bibinfo{person}{Carlos Castillo}, {and}
  \bibinfo{person}{Patrick Meier}.} \bibinfo{year}{2014}\natexlab{}.
\newblock \showarticletitle{Tweetcred: Real-time credibility assessment of
  content on twitter}. In \bibinfo{booktitle}{\emph{International Conference on
  Social Informatics}}. Springer, \bibinfo{pages}{228--243}.
\newblock


\bibitem[\protect\citeauthoryear{Gupta, Lamba, and Kumaraguru}{Gupta
  et~al\mbox{.}}{2013a}]%
        {gupta20131}
\bibfield{author}{\bibinfo{person}{Aditi Gupta}, \bibinfo{person}{Hemank
  Lamba}, {and} \bibinfo{person}{Ponnurangam Kumaraguru}.}
  \bibinfo{year}{2013}\natexlab{a}.
\newblock \showarticletitle{prayforboston: Analyzing fake content on twitter}.
  In \bibinfo{booktitle}{\emph{eCrime Researchers Summit (eCRS), 2013}}. IEEE,
  \bibinfo{pages}{1--12}.
\newblock


\bibitem[\protect\citeauthoryear{Gupta, Lamba, Kumaraguru, and Joshi}{Gupta
  et~al\mbox{.}}{2013b}]%
        {gupta2013faking}
\bibfield{author}{\bibinfo{person}{Aditi Gupta}, \bibinfo{person}{Hemank
  Lamba}, \bibinfo{person}{Ponnurangam Kumaraguru}, {and}
  \bibinfo{person}{Anupam Joshi}.} \bibinfo{year}{2013}\natexlab{b}.
\newblock \showarticletitle{Faking sandy: characterizing and identifying fake
  images on twitter during hurricane sandy}. In
  \bibinfo{booktitle}{\emph{Proceedings of the 22nd international conference on
  World Wide Web}}. ACM, \bibinfo{pages}{729--736}.
\newblock


\bibitem[\protect\citeauthoryear{Hassan, Arslan, Li, and Tremayne}{Hassan
  et~al\mbox{.}}{2017a}]%
        {hassan2017toward}
\bibfield{author}{\bibinfo{person}{Naeemul Hassan}, \bibinfo{person}{Fatma
  Arslan}, \bibinfo{person}{Chengkai Li}, {and} \bibinfo{person}{Mark
  Tremayne}.} \bibinfo{year}{2017}\natexlab{a}.
\newblock \showarticletitle{Toward automated fact-checking: Detecting
  check-worthy factual claims by ClaimBuster}. In
  \bibinfo{booktitle}{\emph{Proceedings of the 23rd ACM SIGKDD International
  Conference on Knowledge Discovery and Data Mining}}. ACM,
  \bibinfo{pages}{1803--1812}.
\newblock


\bibitem[\protect\citeauthoryear{Hassan, Zhang, Arslan, Caraballo, Jimenez,
  Gawsane, Hasan, Joseph, Kulkarni, Nayak, et~al\mbox{.}}{Hassan
  et~al\mbox{.}}{2017b}]%
        {hassan2017claimbuster}
\bibfield{author}{\bibinfo{person}{Naeemul Hassan}, \bibinfo{person}{Gensheng
  Zhang}, \bibinfo{person}{Fatma Arslan}, \bibinfo{person}{Josue Caraballo},
  \bibinfo{person}{Damian Jimenez}, \bibinfo{person}{Siddhant Gawsane},
  \bibinfo{person}{Shohedul Hasan}, \bibinfo{person}{Minumol Joseph},
  \bibinfo{person}{Aaditya Kulkarni}, \bibinfo{person}{Anil~Kumar Nayak},
  {et~al\mbox{.}}} \bibinfo{year}{2017}\natexlab{b}.
\newblock \showarticletitle{ClaimBuster: the first-ever end-to-end
  fact-checking system}.
\newblock \bibinfo{journal}{\emph{Proceedings of the VLDB Endowment}}
  \bibinfo{volume}{10}, \bibinfo{number}{12} (\bibinfo{year}{2017}),
  \bibinfo{pages}{1945--1948}.
\newblock


\bibitem[\protect\citeauthoryear{Herman and Chomsky}{Herman and
  Chomsky}{2010}]%
        {herman2010manufacturing}
\bibfield{author}{\bibinfo{person}{Edward~S Herman} {and} \bibinfo{person}{Noam
  Chomsky}.} \bibinfo{year}{2010}\natexlab{}.
\newblock \bibinfo{booktitle}{\emph{Manufacturing consent: The political
  economy of the mass media}}.
\newblock \bibinfo{publisher}{Random House}.
\newblock


\bibitem[\protect\citeauthoryear{Howell et~al\mbox{.}}{Howell
  et~al\mbox{.}}{2013}]%
        {howell2013digital}
\bibfield{author}{\bibinfo{person}{Lee Howell} {et~al\mbox{.}}}
  \bibinfo{year}{2013}\natexlab{}.
\newblock \showarticletitle{Digital wildfires in a hyperconnected world}.
\newblock \bibinfo{journal}{\emph{WEF Report}}  \bibinfo{volume}{3}
  (\bibinfo{year}{2013}), \bibinfo{pages}{15--94}.
\newblock


\bibitem[\protect\citeauthoryear{Huang, Starbird, Orand, Stanek, and
  Pedersen}{Huang et~al\mbox{.}}{2015}]%
        {huang2015connected}
\bibfield{author}{\bibinfo{person}{Y~Linlin Huang}, \bibinfo{person}{Kate
  Starbird}, \bibinfo{person}{Mania Orand}, \bibinfo{person}{Stephanie~A
  Stanek}, {and} \bibinfo{person}{Heather~T Pedersen}.}
  \bibinfo{year}{2015}\natexlab{}.
\newblock \showarticletitle{Connected through crisis: Emotional proximity and
  the spread of misinformation online}. In
  \bibinfo{booktitle}{\emph{Proceedings of the 18th ACM Conference on Computer
  Supported Cooperative Work \& Social Computing}}. ACM,
  \bibinfo{pages}{969--980}.
\newblock


\bibitem[\protect\citeauthoryear{Hwang, Pearce, and Nanis}{Hwang
  et~al\mbox{.}}{2012}]%
        {hwang2012socialbots}
\bibfield{author}{\bibinfo{person}{Tim Hwang}, \bibinfo{person}{Ian Pearce},
  {and} \bibinfo{person}{Max Nanis}.} \bibinfo{year}{2012}\natexlab{}.
\newblock \showarticletitle{Socialbots: Voices from the fronts}.
\newblock \bibinfo{journal}{\emph{interactions}} \bibinfo{volume}{19},
  \bibinfo{number}{2} (\bibinfo{year}{2012}), \bibinfo{pages}{38--45}.
\newblock


\bibitem[\protect\citeauthoryear{Jack}{Jack}{2017}]%
        {jack2017lexicon}
\bibfield{author}{\bibinfo{person}{Caroline Jack}.}
  \bibinfo{year}{2017}\natexlab{}.
\newblock \showarticletitle{Lexicon of Lies: Terms for Problematic
  Information}.
\newblock \bibinfo{journal}{\emph{Data \& Society}}  \bibinfo{volume}{3}
  (\bibinfo{year}{2017}).
\newblock


\bibitem[\protect\citeauthoryear{Jin, Cao, Zhang, and Luo}{Jin
  et~al\mbox{.}}{2016}]%
        {jin2016news}
\bibfield{author}{\bibinfo{person}{Zhiwei Jin}, \bibinfo{person}{Juan Cao},
  \bibinfo{person}{Yongdong Zhang}, {and} \bibinfo{person}{Jiebo Luo}.}
  \bibinfo{year}{2016}\natexlab{}.
\newblock \showarticletitle{News Verification by Exploiting Conflicting Social
  Viewpoints in Microblogs.}. In \bibinfo{booktitle}{\emph{AAAI}}.
  \bibinfo{pages}{2972--2978}.
\newblock


\bibitem[\protect\citeauthoryear{Jin, Cao, Zhang, Zhou, and Tian}{Jin
  et~al\mbox{.}}{2017}]%
        {jin2017novel}
\bibfield{author}{\bibinfo{person}{Zhiwei Jin}, \bibinfo{person}{Juan Cao},
  \bibinfo{person}{Yongdong Zhang}, \bibinfo{person}{Jianshe Zhou}, {and}
  \bibinfo{person}{Qi Tian}.} \bibinfo{year}{2017}\natexlab{}.
\newblock \showarticletitle{Novel visual and statistical image features for
  microblogs news verification}.
\newblock \bibinfo{journal}{\emph{IEEE transactions on multimedia}}
  \bibinfo{volume}{19}, \bibinfo{number}{3} (\bibinfo{year}{2017}),
  \bibinfo{pages}{598--608}.
\newblock


\bibitem[\protect\citeauthoryear{Karduni, Wesslen, Santhanam, Cho, Volkova,
  Arendt, Shaikh, and Dou}{Karduni et~al\mbox{.}}{2018}]%
        {karduni2018can}
\bibfield{author}{\bibinfo{person}{Alireza Karduni}, \bibinfo{person}{Ryan
  Wesslen}, \bibinfo{person}{Sashank Santhanam}, \bibinfo{person}{Isaac Cho},
  \bibinfo{person}{Svitlana Volkova}, \bibinfo{person}{Dustin Arendt},
  \bibinfo{person}{Samira Shaikh}, {and} \bibinfo{person}{Wenwen Dou}.}
  \bibinfo{year}{2018}\natexlab{}.
\newblock \showarticletitle{Can You Verifi This? Studying Uncertainty and
  Decision-Making About Misinformation using Visual Analytics}. In
  \bibinfo{booktitle}{\emph{International Conference on Web and Social Media
  (ICWSM)}}.
\newblock


\bibitem[\protect\citeauthoryear{Kasra, Shen, and O'Brien}{Kasra
  et~al\mbox{.}}{2018}]%
        {kasra2018seeing}
\bibfield{author}{\bibinfo{person}{Mona Kasra}, \bibinfo{person}{Cuihua Shen},
  {and} \bibinfo{person}{James~F O'Brien}.} \bibinfo{year}{2018}\natexlab{}.
\newblock \showarticletitle{Seeing Is Believing: How People Fail to Identify
  Fake Images on the Web}. In \bibinfo{booktitle}{\emph{Extended Abstracts of
  the 2018 CHI Conference on Human Factors in Computing Systems}}. ACM,
  \bibinfo{pages}{LBW516}.
\newblock


\bibitem[\protect\citeauthoryear{Klein and Wueller}{Klein and Wueller}{2017}]%
        {klein2017fake}
\bibfield{author}{\bibinfo{person}{David Klein} {and} \bibinfo{person}{Joshua
  Wueller}.} \bibinfo{year}{2017}\natexlab{}.
\newblock \showarticletitle{Fake news: A legal perspective}.
\newblock  (\bibinfo{year}{2017}).
\newblock


\bibitem[\protect\citeauthoryear{Koppel and Schler}{Koppel and Schler}{2004}]%
        {koppel2004authorship}
\bibfield{author}{\bibinfo{person}{Moshe Koppel} {and}
  \bibinfo{person}{Jonathan Schler}.} \bibinfo{year}{2004}\natexlab{}.
\newblock \showarticletitle{Authorship verification as a one-class
  classification problem}. In \bibinfo{booktitle}{\emph{Proceedings of the
  twenty-first international conference on Machine learning}}. ACM,
  \bibinfo{pages}{62}.
\newblock


\bibitem[\protect\citeauthoryear{Lakoff}{Lakoff}{2010}]%
        {lakoff2010moral}
\bibfield{author}{\bibinfo{person}{George Lakoff}.}
  \bibinfo{year}{2010}\natexlab{}.
\newblock \bibinfo{booktitle}{\emph{Moral politics: How liberals and
  conservatives think}}.
\newblock \bibinfo{publisher}{University of Chicago Press}.
\newblock


\bibitem[\protect\citeauthoryear{Lazer, Baum, Grinberg, Friedland, Joseph,
  Hobbs, and Mattsson}{Lazer et~al\mbox{.}}{2017}]%
        {lazer2017combating}
\bibfield{author}{\bibinfo{person}{David Lazer}, \bibinfo{person}{Matthew
  Baum}, \bibinfo{person}{Nir Grinberg}, \bibinfo{person}{Lisa Friedland},
  \bibinfo{person}{Kenneth Joseph}, \bibinfo{person}{Will Hobbs}, {and}
  \bibinfo{person}{Carolina Mattsson}.} \bibinfo{year}{2017}\natexlab{}.
\newblock \showarticletitle{Combating fake news: An agenda for research and
  action}.
\newblock \bibinfo{journal}{\emph{Harvard Kennedy School, Shorenstein Center on
  Media, Politics and Public Policy}}  \bibinfo{volume}{2}
  (\bibinfo{year}{2017}).
\newblock


\bibitem[\protect\citeauthoryear{Lazer, Baum, Benkler, Berinsky, Greenhill,
  Menczer, Metzger, Nyhan, Pennycook, Rothschild, et~al\mbox{.}}{Lazer
  et~al\mbox{.}}{2018}]%
        {lazer2018science}
\bibfield{author}{\bibinfo{person}{David~MJ Lazer}, \bibinfo{person}{Matthew~A
  Baum}, \bibinfo{person}{Yochai Benkler}, \bibinfo{person}{Adam~J Berinsky},
  \bibinfo{person}{Kelly~M Greenhill}, \bibinfo{person}{Filippo Menczer},
  \bibinfo{person}{Miriam~J Metzger}, \bibinfo{person}{Brendan Nyhan},
  \bibinfo{person}{Gordon Pennycook}, \bibinfo{person}{David Rothschild},
  {et~al\mbox{.}}} \bibinfo{year}{2018}\natexlab{}.
\newblock \showarticletitle{The science of fake news}.
\newblock \bibinfo{journal}{\emph{Science}} \bibinfo{volume}{359},
  \bibinfo{number}{6380} (\bibinfo{year}{2018}), \bibinfo{pages}{1094--1096}.
\newblock


\bibitem[\protect\citeauthoryear{Ma, Gao, Wei, Lu, and Wong}{Ma
  et~al\mbox{.}}{2015}]%
        {ma2015detect}
\bibfield{author}{\bibinfo{person}{Jing Ma}, \bibinfo{person}{Wei Gao},
  \bibinfo{person}{Zhongyu Wei}, \bibinfo{person}{Yueming Lu}, {and}
  \bibinfo{person}{Kam-Fai Wong}.} \bibinfo{year}{2015}\natexlab{}.
\newblock \showarticletitle{Detect rumors using time series of social context
  information on microblogging websites}. In
  \bibinfo{booktitle}{\emph{Proceedings of the 24th ACM International on
  Conference on Information and Knowledge Management}}. ACM,
  \bibinfo{pages}{1751--1754}.
\newblock


\bibitem[\protect\citeauthoryear{Magdy and Wanas}{Magdy and Wanas}{2010}]%
        {magdy2010web}
\bibfield{author}{\bibinfo{person}{Amr Magdy} {and} \bibinfo{person}{Nayer
  Wanas}.} \bibinfo{year}{2010}\natexlab{}.
\newblock \showarticletitle{Web-based statistical fact checking of textual
  documents}. In \bibinfo{booktitle}{\emph{Proceedings of the 2nd international
  workshop on Search and mining user-generated contents}}. ACM,
  \bibinfo{pages}{103--110}.
\newblock


\bibitem[\protect\citeauthoryear{Mallonee}{Mallonee}{2017}]%
        {mallonee_2017}
\bibfield{author}{\bibinfo{person}{Laura Mallonee}.}
  \bibinfo{year}{2017}\natexlab{}.
\newblock \bibinfo{title}{How Photos Fuel the Spread of Fake News}.
\newblock
\newblock
\urldef\tempurl%
\url{https://www.wired.com/2016/12/photos-fuel-spread-fake-news/}
\showURL{%
\tempurl}


\bibitem[\protect\citeauthoryear{Marcus}{Marcus}{1992}]%
        {marcus1992mesoamerican}
\bibfield{author}{\bibinfo{person}{Joyce Marcus}.}
  \bibinfo{year}{1992}\natexlab{}.
\newblock \bibinfo{booktitle}{\emph{Mesoamerican writing systems: Propaganda,
  myth, and history in four ancient civilizations}}.
\newblock \bibinfo{publisher}{Princeton University Press Princeton}.
\newblock


\bibitem[\protect\citeauthoryear{Mele, Lazer, Baum, Grinberg, Friedland,
  Joseph, Hobbs, and Mattsson}{Mele et~al\mbox{.}}{2017}]%
        {mele2017combating}
\bibfield{author}{\bibinfo{person}{Nicco Mele}, \bibinfo{person}{David Lazer},
  \bibinfo{person}{Matthew Baum}, \bibinfo{person}{Nir Grinberg},
  \bibinfo{person}{Lisa Friedland}, \bibinfo{person}{Kenneth Joseph},
  \bibinfo{person}{Will Hobbs}, {and} \bibinfo{person}{Carolina Mattsson}.}
  \bibinfo{year}{2017}\natexlab{}.
\newblock \bibinfo{title}{Combating fake news: An agenda for research and
  action}.
\newblock
\newblock


\bibitem[\protect\citeauthoryear{Messaris and Abraham}{Messaris and
  Abraham}{2001}]%
        {messaris2001role}
\bibfield{author}{\bibinfo{person}{Paul Messaris} {and} \bibinfo{person}{Linus
  Abraham}.} \bibinfo{year}{2001}\natexlab{}.
\newblock \showarticletitle{The role of images in framing news stories}.
\newblock \bibinfo{journal}{\emph{Framing public life: Perspectives on media
  and our understanding of the social world}} (\bibinfo{year}{2001}),
  \bibinfo{pages}{215--226}.
\newblock


\bibitem[\protect\citeauthoryear{Mynatt, Doherty, and Tweney}{Mynatt
  et~al\mbox{.}}{1977}]%
        {mynatt1977confirmation}
\bibfield{author}{\bibinfo{person}{Clifford~R Mynatt},
  \bibinfo{person}{Michael~E Doherty}, {and} \bibinfo{person}{Ryan~D Tweney}.}
  \bibinfo{year}{1977}\natexlab{}.
\newblock \showarticletitle{Confirmation bias in a simulated research
  environment: An experimental study of scientific inference}.
\newblock \bibinfo{journal}{\emph{The quarterly journal of experimental
  psychology}} \bibinfo{volume}{29}, \bibinfo{number}{1}
  (\bibinfo{year}{1977}), \bibinfo{pages}{85--95}.
\newblock


\bibitem[\protect\citeauthoryear{Narwal, Salih, Lopez, Ortega, O'Donovan,
  H{\"o}llerer, and Savage}{Narwal et~al\mbox{.}}{2017}]%
        {narwal2017automated}
\bibfield{author}{\bibinfo{person}{Vishwajeet Narwal},
  \bibinfo{person}{Mohamed~Hashim Salih}, \bibinfo{person}{Jose~Angel Lopez},
  \bibinfo{person}{Angel Ortega}, \bibinfo{person}{John O'Donovan},
  \bibinfo{person}{Tobias H{\"o}llerer}, {and} \bibinfo{person}{Saiph Savage}.}
  \bibinfo{year}{2017}\natexlab{}.
\newblock \showarticletitle{Automated assistants to identify and prompt action
  on visual news bias}. In \bibinfo{booktitle}{\emph{Proceedings of the 2017
  CHI Conference Extended Abstracts on Human Factors in Computing Systems}}.
  ACM, \bibinfo{pages}{2796--2801}.
\newblock


\bibitem[\protect\citeauthoryear{Newman, Fletcher, Kalogeropoulos, Levy, and
  Nielsen}{Newman et~al\mbox{.}}{2017}]%
        {newman2017reuters}
\bibfield{author}{\bibinfo{person}{Nic Newman}, \bibinfo{person}{Richard
  Fletcher}, \bibinfo{person}{Antonis Kalogeropoulos},
  \bibinfo{person}{David~AL Levy}, {and} \bibinfo{person}{Rasmus~Kleis
  Nielsen}.} \bibinfo{year}{2017}\natexlab{}.
\newblock \showarticletitle{Reuters Institute digital news report 2017}.
\newblock  (\bibinfo{year}{2017}).
\newblock


\bibitem[\protect\citeauthoryear{Nyhan and Reifler}{Nyhan and Reifler}{2010}]%
        {nyhan2010corrections}
\bibfield{author}{\bibinfo{person}{Brendan Nyhan} {and} \bibinfo{person}{Jason
  Reifler}.} \bibinfo{year}{2010}\natexlab{}.
\newblock \showarticletitle{When corrections fail: The persistence of political
  misperceptions}.
\newblock \bibinfo{journal}{\emph{Political Behavior}} \bibinfo{volume}{32},
  \bibinfo{number}{2} (\bibinfo{year}{2010}), \bibinfo{pages}{303--330}.
\newblock


\bibitem[\protect\citeauthoryear{Oraby, Reed, Compton, Riloff, Walker, and
  Whittaker}{Oraby et~al\mbox{.}}{2017}]%
        {oraby2017and}
\bibfield{author}{\bibinfo{person}{Shereen Oraby}, \bibinfo{person}{Lena Reed},
  \bibinfo{person}{Ryan Compton}, \bibinfo{person}{Ellen Riloff},
  \bibinfo{person}{Marilyn Walker}, {and} \bibinfo{person}{Steve Whittaker}.}
  \bibinfo{year}{2017}\natexlab{}.
\newblock \showarticletitle{And that's a fact: Distinguishing factual and
  emotional argumentation in online dialogue}.
\newblock \bibinfo{journal}{\emph{arXiv preprint arXiv:1709.05295}}
  (\bibinfo{year}{2017}).
\newblock


\bibitem[\protect\citeauthoryear{Ott, Choi, Cardie, and Hancock}{Ott
  et~al\mbox{.}}{2011}]%
        {ott2011finding}
\bibfield{author}{\bibinfo{person}{Myle Ott}, \bibinfo{person}{Yejin Choi},
  \bibinfo{person}{Claire Cardie}, {and} \bibinfo{person}{Jeffrey~T Hancock}.}
  \bibinfo{year}{2011}\natexlab{}.
\newblock \showarticletitle{Finding deceptive opinion spam by any stretch of
  the imagination}. In \bibinfo{booktitle}{\emph{Proceedings of the 49th Annual
  Meeting of the Association for Computational Linguistics: Human Language
  Technologies-Volume 1}}. Association for Computational Linguistics,
  \bibinfo{pages}{309--319}.
\newblock


\bibitem[\protect\citeauthoryear{Pariser}{Pariser}{2011}]%
        {pariser2011filter}
\bibfield{author}{\bibinfo{person}{Eli Pariser}.}
  \bibinfo{year}{2011}\natexlab{}.
\newblock \bibinfo{booktitle}{\emph{The filter bubble: What the Internet is
  hiding from you}}.
\newblock \bibinfo{publisher}{Penguin UK}.
\newblock


\bibitem[\protect\citeauthoryear{Pennebaker, Boyd, Jordan, and
  Blackburn}{Pennebaker et~al\mbox{.}}{2015}]%
        {pennebaker2015development}
\bibfield{author}{\bibinfo{person}{James~W Pennebaker}, \bibinfo{person}{Ryan~L
  Boyd}, \bibinfo{person}{Kayla Jordan}, {and} \bibinfo{person}{Kate
  Blackburn}.} \bibinfo{year}{2015}\natexlab{}.
\newblock \bibinfo{booktitle}{\emph{The development and psychometric properties
  of LIWC2015}}.
\newblock \bibinfo{type}{{T}echnical {R}eport}.
\newblock


\bibitem[\protect\citeauthoryear{Pennycook, Cannon, and Rand}{Pennycook
  et~al\mbox{.}}{2018}]%
        {pennycook2018prior}
\bibfield{author}{\bibinfo{person}{Gordon Pennycook}, \bibinfo{person}{Tyrone
  Cannon}, {and} \bibinfo{person}{David~G Rand}.}
  \bibinfo{year}{2018}\natexlab{}.
\newblock \showarticletitle{Prior exposure increases perceived accuracy of fake
  news}.
\newblock  (\bibinfo{year}{2018}).
\newblock


\bibitem[\protect\citeauthoryear{Pennycook, Cannon, and Rand}{Pennycook
  et~al\mbox{.}}{2017}]%
        {pennycook2017prior}
\bibfield{author}{\bibinfo{person}{Gordon Pennycook}, \bibinfo{person}{Tyrone~D
  Cannon}, {and} \bibinfo{person}{David~G Rand}.}
  \bibinfo{year}{2017}\natexlab{}.
\newblock \showarticletitle{Implausibility and Illusory Truth: Prior Exposure
  Increases Perceived Accuracy of Fake News but Has No Effect on Entirely
  Implausible Statements}.
\newblock \bibinfo{journal}{\emph{Available at SSRN}} (\bibinfo{year}{2017}).
\newblock


\bibitem[\protect\citeauthoryear{Pennycook and Rand}{Pennycook and
  Rand}{2018a}]%
        {pennycook2018cognitive}
\bibfield{author}{\bibinfo{person}{Gordon Pennycook} {and}
  \bibinfo{person}{David~G Rand}.} \bibinfo{year}{2018}\natexlab{a}.
\newblock \showarticletitle{Cognitive Reflection and the 2016 US Presidential
  Election}.
\newblock \bibinfo{journal}{\emph{Forthcoming in Personality and Social
  Psychology Bulletin}} (\bibinfo{year}{2018}).
\newblock


\bibitem[\protect\citeauthoryear{Pennycook and Rand}{Pennycook and
  Rand}{2018b}]%
        {pennycook2018falls}
\bibfield{author}{\bibinfo{person}{Gordon Pennycook} {and}
  \bibinfo{person}{David~G Rand}.} \bibinfo{year}{2018}\natexlab{b}.
\newblock \showarticletitle{Who falls for fake news? The roles of bullshit
  receptivity, overclaiming, familiarity, and analytic thinking}.
\newblock  (\bibinfo{year}{2018}).
\newblock


\bibitem[\protect\citeauthoryear{Perlmutter}{Perlmutter}{1968}]%
        {perlmutter1968deep}
\bibfield{author}{\bibinfo{person}{David~M Perlmutter}.}
  \bibinfo{year}{1968}\natexlab{}.
\newblock \emph{\bibinfo{title}{Deep and surface structure constraints in
  syntax.}}
\newblock \bibinfo{thesistype}{Ph.D. Dissertation}.
  \bibinfo{school}{Massachusetts Institute of Technology}.
\newblock


\bibitem[\protect\citeauthoryear{Peters}{Peters}{[n. d.]}]%
        {Wielding48:online}
\bibfield{author}{\bibinfo{person}{Jeremy~W. Peters}.} \bibinfo{year}{[n.
  d.]}\natexlab{}.
\newblock \bibinfo{title}{Wielding Claims of ‘Fake News,’ Conservatives
  Take Aim at Mainstream Media - The New York Times}.
\newblock
  \bibinfo{howpublished}{\url{https://www.nytimes.com/2016/12/25/us/politics/fake-news-claims-conservatives-mainstream-media-.html?_r=0}}.
\newblock
\newblock
\shownote{(Accessed on 12/01/2018).}


\bibitem[\protect\citeauthoryear{Potthast, Kiesel, Reinartz, Bevendorff, and
  Stein}{Potthast et~al\mbox{.}}{2017}]%
        {potthast2017stylometric}
\bibfield{author}{\bibinfo{person}{Martin Potthast}, \bibinfo{person}{Johannes
  Kiesel}, \bibinfo{person}{Kevin Reinartz}, \bibinfo{person}{Janek
  Bevendorff}, {and} \bibinfo{person}{Benno Stein}.}
  \bibinfo{year}{2017}\natexlab{}.
\newblock \showarticletitle{A stylometric inquiry into hyperpartisan and fake
  news}.
\newblock \bibinfo{journal}{\emph{arXiv preprint arXiv:1702.05638}}
  (\bibinfo{year}{2017}).
\newblock


\bibitem[\protect\citeauthoryear{Pourghomi, Safieddine, Masri, and
  Dordevic}{Pourghomi et~al\mbox{.}}{2017}]%
        {pourghomi2017stop}
\bibfield{author}{\bibinfo{person}{Pardis Pourghomi}, \bibinfo{person}{Fadi
  Safieddine}, \bibinfo{person}{Wassim Masri}, {and} \bibinfo{person}{Milan
  Dordevic}.} \bibinfo{year}{2017}\natexlab{}.
\newblock \showarticletitle{How to stop spread of misinformation on social
  media: Facebook plans vs. right-click authenticate approach}. In
  \bibinfo{booktitle}{\emph{Engineering \& MIS (ICEMIS), 2017 International
  Conference on}}. IEEE, \bibinfo{pages}{1--8}.
\newblock


\bibitem[\protect\citeauthoryear{Ratkiewicz, Conover, Meiss, Gon{\c{c}}alves,
  Flammini, and Menczer}{Ratkiewicz et~al\mbox{.}}{2011}]%
        {ratkiewicz2011detecting}
\bibfield{author}{\bibinfo{person}{Jacob Ratkiewicz}, \bibinfo{person}{Michael
  Conover}, \bibinfo{person}{Mark~R Meiss}, \bibinfo{person}{Bruno
  Gon{\c{c}}alves}, \bibinfo{person}{Alessandro Flammini}, {and}
  \bibinfo{person}{Filippo Menczer}.} \bibinfo{year}{2011}\natexlab{}.
\newblock \showarticletitle{Detecting and tracking political abuse in social
  media.}
\newblock \bibinfo{journal}{\emph{ICWSM}}  \bibinfo{volume}{11}
  (\bibinfo{year}{2011}), \bibinfo{pages}{297--304}.
\newblock


\bibitem[\protect\citeauthoryear{Resnick, Carton, Park, Shen, and
  Zeffer}{Resnick et~al\mbox{.}}{2014}]%
        {resnick2014rumorlens}
\bibfield{author}{\bibinfo{person}{Paul Resnick}, \bibinfo{person}{Samuel
  Carton}, \bibinfo{person}{Souneil Park}, \bibinfo{person}{Yuncheng Shen},
  {and} \bibinfo{person}{Nicole Zeffer}.} \bibinfo{year}{2014}\natexlab{}.
\newblock \showarticletitle{Rumorlens: A system for analyzing the impact of
  rumors and corrections in social media}. In \bibinfo{booktitle}{\emph{Proc.
  Computational Journalism Conference}}.
\newblock


\bibitem[\protect\citeauthoryear{Robb}{Robb}{[n. d.]}]%
        {Pizzagat59:online}
\bibfield{author}{\bibinfo{person}{Amanda Robb}.} \bibinfo{year}{[n.
  d.]}\natexlab{}.
\newblock \bibinfo{title}{Pizzagate: Anatomy of a Fake News Scandal – Rolling
  Stone}.
\newblock
  \bibinfo{howpublished}{\url{https://www.rollingstone.com/politics/politics-news/anatomy-of-a-fake-news-scandal-125877/}}.
\newblock
\newblock
\shownote{(Accessed on 12/02/2018).}


\bibitem[\protect\citeauthoryear{Rubin, Conroy, Chen, and Cornwell}{Rubin
  et~al\mbox{.}}{2016}]%
        {rubin2016fake}
\bibfield{author}{\bibinfo{person}{Victoria Rubin}, \bibinfo{person}{Niall
  Conroy}, \bibinfo{person}{Yimin Chen}, {and} \bibinfo{person}{Sarah
  Cornwell}.} \bibinfo{year}{2016}\natexlab{}.
\newblock \showarticletitle{Fake news or truth? using satirical cues to detect
  potentially misleading news}. In \bibinfo{booktitle}{\emph{Proceedings of the
  Second Workshop on Computational Approaches to Deception Detection}}.
  \bibinfo{pages}{7--17}.
\newblock


\bibitem[\protect\citeauthoryear{Rubin, Chen, and Conroy}{Rubin
  et~al\mbox{.}}{2015}]%
        {rubin2015deception}
\bibfield{author}{\bibinfo{person}{Victoria~L Rubin}, \bibinfo{person}{Yimin
  Chen}, {and} \bibinfo{person}{Niall~J Conroy}.}
  \bibinfo{year}{2015}\natexlab{}.
\newblock \showarticletitle{Deception detection for news: three types of
  fakes}. In \bibinfo{booktitle}{\emph{Proceedings of the 78th ASIS\&T Annual
  Meeting: Information Science with Impact: Research in and for the
  Community}}. American Society for Information Science, \bibinfo{pages}{83}.
\newblock


\bibitem[\protect\citeauthoryear{Schwartz}{Schwartz}{[n. d.]}]%
        {Trump’s‘26:online}
\bibfield{author}{\bibinfo{person}{Jason Schwartz}.} \bibinfo{year}{[n.
  d.]}\natexlab{}.
\newblock \bibinfo{title}{Trump’s ‘fake news’ rhetoric crops up around
  the globe – POLITICO}.
\newblock
  \bibinfo{howpublished}{\url{https://www.politico.eu/blogs/on-media/2018/07/donald-trump-fake-news-rhetoric-crops-up-around-the-globe-media-social-media-foreign-affairs/}}.
\newblock
\newblock
\shownote{(Accessed on 12/01/2018).}


\bibitem[\protect\citeauthoryear{Shao, Ciampaglia, Flammini, and Menczer}{Shao
  et~al\mbox{.}}{2016}]%
        {shao2016hoaxy}
\bibfield{author}{\bibinfo{person}{Chengcheng Shao},
  \bibinfo{person}{Giovanni~Luca Ciampaglia}, \bibinfo{person}{Alessandro
  Flammini}, {and} \bibinfo{person}{Filippo Menczer}.}
  \bibinfo{year}{2016}\natexlab{}.
\newblock \showarticletitle{Hoaxy: A platform for tracking online
  misinformation}. In \bibinfo{booktitle}{\emph{Proceedings of the 25th
  international conference companion on world wide web}}. International World
  Wide Web Conferences Steering Committee, \bibinfo{pages}{745--750}.
\newblock


\bibitem[\protect\citeauthoryear{Shao, Ciampaglia, Varol, Flammini, and
  Menczer}{Shao et~al\mbox{.}}{2017}]%
        {shao2017spread}
\bibfield{author}{\bibinfo{person}{Chengcheng Shao},
  \bibinfo{person}{Giovanni~Luca Ciampaglia}, \bibinfo{person}{Onur Varol},
  \bibinfo{person}{Alessandro Flammini}, {and} \bibinfo{person}{Filippo
  Menczer}.} \bibinfo{year}{2017}\natexlab{}.
\newblock \showarticletitle{The spread of fake news by social bots}.
\newblock \bibinfo{journal}{\emph{arXiv preprint arXiv:1707.07592}}
  (\bibinfo{year}{2017}), \bibinfo{pages}{96--104}.
\newblock


\bibitem[\protect\citeauthoryear{Shu, Sliva, Wang, Tang, and Liu}{Shu
  et~al\mbox{.}}{2017}]%
        {shu2017fake}
\bibfield{author}{\bibinfo{person}{Kai Shu}, \bibinfo{person}{Amy Sliva},
  \bibinfo{person}{Suhang Wang}, \bibinfo{person}{Jiliang Tang}, {and}
  \bibinfo{person}{Huan Liu}.} \bibinfo{year}{2017}\natexlab{}.
\newblock \showarticletitle{Fake news detection on social media: A data mining
  perspective}.
\newblock \bibinfo{journal}{\emph{ACM SIGKDD Explorations Newsletter}}
  \bibinfo{volume}{19}, \bibinfo{number}{1} (\bibinfo{year}{2017}),
  \bibinfo{pages}{22--36}.
\newblock


\bibitem[\protect\citeauthoryear{Sloman and Fernbach}{Sloman and
  Fernbach}{2018}]%
        {sloman2018knowledge}
\bibfield{author}{\bibinfo{person}{Steven Sloman} {and} \bibinfo{person}{Philip
  Fernbach}.} \bibinfo{year}{2018}\natexlab{}.
\newblock \bibinfo{booktitle}{\emph{The knowledge illusion: Why we never think
  alone}}.
\newblock \bibinfo{publisher}{Penguin}.
\newblock


\bibitem[\protect\citeauthoryear{Spohr}{Spohr}{2017}]%
        {spohr2017fake}
\bibfield{author}{\bibinfo{person}{Dominic Spohr}.}
  \bibinfo{year}{2017}\natexlab{}.
\newblock \showarticletitle{Fake news and ideological polarization: Filter
  bubbles and selective exposure on social media}.
\newblock \bibinfo{journal}{\emph{Business Information Review}}
  \bibinfo{volume}{34}, \bibinfo{number}{3} (\bibinfo{year}{2017}),
  \bibinfo{pages}{150--160}.
\newblock


\bibitem[\protect\citeauthoryear{Starbird}{Starbird}{2017}]%
        {starbird2017examining}
\bibfield{author}{\bibinfo{person}{Kate Starbird}.}
  \bibinfo{year}{2017}\natexlab{}.
\newblock \showarticletitle{Examining the Alternative Media Ecosystem Through
  the Production of Alternative Narratives of Mass Shooting Events on
  Twitter.}. In \bibinfo{booktitle}{\emph{ICWSM}}. \bibinfo{pages}{230--239}.
\newblock


\bibitem[\protect\citeauthoryear{Sullivan}{Sullivan}{[n. d.]}]%
        {Whythete18:online}
\bibfield{author}{\bibinfo{person}{Margaret Sullivan}.} \bibinfo{year}{[n.
  d.]}\natexlab{}.
\newblock \bibinfo{title}{Why the term 'fake news' should be retired in 2018 -
  The Washington Post}.
\newblock
  \bibinfo{howpublished}{\url{https://www.washingtonpost.com/lifestyle/style/its-time-to-retire-the-tainted-term-fake-news/2017/01/06/a5a7516c-d375-11e6-945a-76f69a399dd5_story.html?utm_term=.902df60602f3}}.
\newblock
\newblock
\shownote{(Accessed on 12/01/2018).}


\bibitem[\protect\citeauthoryear{Sunstein}{Sunstein}{2001}]%
        {sunstein2001echo}
\bibfield{author}{\bibinfo{person}{Cass~R Sunstein}.}
  \bibinfo{year}{2001}\natexlab{}.
\newblock \bibinfo{booktitle}{\emph{Echo chambers: Bush v. Gore, impeachment,
  and beyond}}.
\newblock \bibinfo{publisher}{Princeton University Press Princeton, NJ}.
\newblock


\bibitem[\protect\citeauthoryear{Swire, Ecker, and Lewandowsky}{Swire
  et~al\mbox{.}}{2017}]%
        {swire2017role}
\bibfield{author}{\bibinfo{person}{Briony Swire}, \bibinfo{person}{Ullrich~KH
  Ecker}, {and} \bibinfo{person}{Stephan Lewandowsky}.}
  \bibinfo{year}{2017}\natexlab{}.
\newblock \showarticletitle{The role of familiarity in correcting inaccurate
  information.}
\newblock \bibinfo{journal}{\emph{Journal of experimental psychology: learning,
  memory, and cognition}} \bibinfo{volume}{43}, \bibinfo{number}{12}
  (\bibinfo{year}{2017}), \bibinfo{pages}{1948}.
\newblock


\bibitem[\protect\citeauthoryear{Tambini}{Tambini}{2017}]%
        {tambini2017fake}
\bibfield{author}{\bibinfo{person}{Damian Tambini}.}
  \bibinfo{year}{2017}\natexlab{}.
\newblock \showarticletitle{Fake news: public policy responses}.
\newblock  (\bibinfo{year}{2017}).
\newblock


\bibitem[\protect\citeauthoryear{Tamibini}{Tamibini}{2017}]%
        {tamibini2017advertising}
\bibfield{author}{\bibinfo{person}{Damian Tamibini}.}
  \bibinfo{year}{2017}\natexlab{}.
\newblock \showarticletitle{How advertising fuels fake news}.
\newblock \bibinfo{journal}{\emph{Media Policy Blog}} (\bibinfo{year}{2017}).
\newblock


\bibitem[\protect\citeauthoryear{Tandoc~Jr, Lim, and Ling}{Tandoc~Jr
  et~al\mbox{.}}{2018}]%
        {tandoc2018defining}
\bibfield{author}{\bibinfo{person}{Edson~C Tandoc~Jr},
  \bibinfo{person}{Zheng~Wei Lim}, {and} \bibinfo{person}{Richard Ling}.}
  \bibinfo{year}{2018}\natexlab{}.
\newblock \showarticletitle{Defining “fake news” A typology of scholarly
  definitions}.
\newblock \bibinfo{journal}{\emph{Digital Journalism}} \bibinfo{volume}{6},
  \bibinfo{number}{2} (\bibinfo{year}{2018}), \bibinfo{pages}{137--153}.
\newblock


\bibitem[\protect\citeauthoryear{Taylor}{Taylor}{2013}]%
        {taylor2013munitions}
\bibfield{author}{\bibinfo{person}{Philip~M Taylor}.}
  \bibinfo{year}{2013}\natexlab{}.
\newblock \showarticletitle{Munitions of the mind: A history of propaganda from
  the ancient world to the present era}.
\newblock  (\bibinfo{year}{2013}).
\newblock


\bibitem[\protect\citeauthoryear{Tversky and Kahneman}{Tversky and
  Kahneman}{1973}]%
        {tversky1973availability}
\bibfield{author}{\bibinfo{person}{Amos Tversky} {and} \bibinfo{person}{Daniel
  Kahneman}.} \bibinfo{year}{1973}\natexlab{}.
\newblock \showarticletitle{Availability: A heuristic for judging frequency and
  probability}.
\newblock \bibinfo{journal}{\emph{Cognitive psychology}} \bibinfo{volume}{5},
  \bibinfo{number}{2} (\bibinfo{year}{1973}), \bibinfo{pages}{207--232}.
\newblock


\bibitem[\protect\citeauthoryear{Vargo, Guo, and Amazeen}{Vargo
  et~al\mbox{.}}{2017}]%
        {vargo2017agenda}
\bibfield{author}{\bibinfo{person}{Chris~J Vargo}, \bibinfo{person}{Lei Guo},
  {and} \bibinfo{person}{Michelle~A Amazeen}.} \bibinfo{year}{2017}\natexlab{}.
\newblock \showarticletitle{The agenda-setting power of fake news: A big data
  analysis of the online media landscape from 2014 to 2016}.
\newblock \bibinfo{journal}{\emph{new media \& society}}
  (\bibinfo{year}{2017}), \bibinfo{pages}{1461444817712086}.
\newblock


\bibitem[\protect\citeauthoryear{Vargo, Guo, and Amazeen}{Vargo
  et~al\mbox{.}}{2018}]%
        {vargo2018agenda}
\bibfield{author}{\bibinfo{person}{Chris~J Vargo}, \bibinfo{person}{Lei Guo},
  {and} \bibinfo{person}{Michelle~A Amazeen}.} \bibinfo{year}{2018}\natexlab{}.
\newblock \showarticletitle{The agenda-setting power of fake news: A big data
  analysis of the online media landscape from 2014 to 2016}.
\newblock \bibinfo{journal}{\emph{new media \& society}} \bibinfo{volume}{20},
  \bibinfo{number}{5} (\bibinfo{year}{2018}), \bibinfo{pages}{2028--2049}.
\newblock


\bibitem[\protect\citeauthoryear{Viswanath, Post, Gummadi, and
  Mislove}{Viswanath et~al\mbox{.}}{2011}]%
        {viswanath2011analysis}
\bibfield{author}{\bibinfo{person}{Bimal Viswanath}, \bibinfo{person}{Ansley
  Post}, \bibinfo{person}{Krishna~P Gummadi}, {and} \bibinfo{person}{Alan
  Mislove}.} \bibinfo{year}{2011}\natexlab{}.
\newblock \showarticletitle{An analysis of social network-based sybil
  defenses}.
\newblock \bibinfo{journal}{\emph{ACM SIGCOMM Computer Communication Review}}
  \bibinfo{volume}{41}, \bibinfo{number}{4} (\bibinfo{year}{2011}),
  \bibinfo{pages}{363--374}.
\newblock


\bibitem[\protect\citeauthoryear{Volkova, Shaffer, Jang, and Hodas}{Volkova
  et~al\mbox{.}}{2017}]%
        {volkova2017separating}
\bibfield{author}{\bibinfo{person}{Svitlana Volkova}, \bibinfo{person}{Kyle
  Shaffer}, \bibinfo{person}{Jin~Yea Jang}, {and} \bibinfo{person}{Nathan
  Hodas}.} \bibinfo{year}{2017}\natexlab{}.
\newblock \showarticletitle{Separating facts from fiction: Linguistic models to
  classify suspicious and trusted news posts on twitter}. In
  \bibinfo{booktitle}{\emph{Proceedings of the 55th Annual Meeting of the
  Association for Computational Linguistics (Volume 2: Short Papers)}},
  Vol.~\bibinfo{volume}{2}. \bibinfo{pages}{647--653}.
\newblock


\bibitem[\protect\citeauthoryear{Vrij, Mann, Kristen, and Fisher}{Vrij
  et~al\mbox{.}}{2007}]%
        {vrij2007cues}
\bibfield{author}{\bibinfo{person}{Aldert Vrij}, \bibinfo{person}{Samantha
  Mann}, \bibinfo{person}{Susanne Kristen}, {and} \bibinfo{person}{Ronald~P
  Fisher}.} \bibinfo{year}{2007}\natexlab{}.
\newblock \showarticletitle{Cues to deception and ability to detect lies as a
  function of police interview styles}.
\newblock \bibinfo{journal}{\emph{Law and human behavior}}
  \bibinfo{volume}{31}, \bibinfo{number}{5} (\bibinfo{year}{2007}),
  \bibinfo{pages}{499--518}.
\newblock


\bibitem[\protect\citeauthoryear{Wang, Mohanlal, Wilson, Wang, Metzger, Zheng,
  and Zhao}{Wang et~al\mbox{.}}{2012}]%
        {wang2012social}
\bibfield{author}{\bibinfo{person}{Gang Wang}, \bibinfo{person}{Manish
  Mohanlal}, \bibinfo{person}{Christo Wilson}, \bibinfo{person}{Xiao Wang},
  \bibinfo{person}{Miriam Metzger}, \bibinfo{person}{Haitao Zheng}, {and}
  \bibinfo{person}{Ben~Y Zhao}.} \bibinfo{year}{2012}\natexlab{}.
\newblock \showarticletitle{Social turing tests: Crowdsourcing sybil
  detection}.
\newblock \bibinfo{journal}{\emph{arXiv preprint arXiv:1205.3856}}
  (\bibinfo{year}{2012}).
\newblock


\bibitem[\protect\citeauthoryear{Wang}{Wang}{2017}]%
        {wang2017liar}
\bibfield{author}{\bibinfo{person}{William~Yang Wang}.}
  \bibinfo{year}{2017}\natexlab{}.
\newblock \showarticletitle{" liar, liar pants on fire": A new benchmark
  dataset for fake news detection}.
\newblock \bibinfo{journal}{\emph{arXiv preprint arXiv:1705.00648}}
  (\bibinfo{year}{2017}).
\newblock


\bibitem[\protect\citeauthoryear{Wardle}{Wardle}{[n. d.]}]%
        {Fakenews76:online}
\bibfield{author}{\bibinfo{person}{Claire Wardle}.} \bibinfo{year}{[n.
  d.]}\natexlab{}.
\newblock \bibinfo{title}{Fake news. It's complicated.}
\newblock
  \bibinfo{howpublished}{\url{https://firstdraftnews.org/fake-news-complicated/}}.
\newblock
\newblock
\shownote{(Accessed on 12/01/2018).}


\bibitem[\protect\citeauthoryear{Wardle and Derakhshan}{Wardle and
  Derakhshan}{2017}]%
        {wardle2017information}
\bibfield{author}{\bibinfo{person}{Claire Wardle} {and}
  \bibinfo{person}{Hossein Derakhshan}.} \bibinfo{year}{2017}\natexlab{}.
\newblock \showarticletitle{Information Disorder: Toward an interdisciplinary
  framework for research and policymaking}.
\newblock \bibinfo{journal}{\emph{Council of Europe report, DGI (2017)}}
  \bibinfo{volume}{9} (\bibinfo{year}{2017}).
\newblock


\bibitem[\protect\citeauthoryear{Waterson, Esposito, and Sanusi}{Waterson
  et~al\mbox{.}}{[n. d.]}]%
        {HereIsAl84:online}
\bibfield{author}{\bibinfo{person}{Jim Waterson}, \bibinfo{person}{Brad
  Esposito}, {and} \bibinfo{person}{Victoria Sanusi}.} \bibinfo{year}{[n.
  d.]}\natexlab{}.
\newblock \bibinfo{title}{Here Is All The Fake News About The Manchester Terror
  Attack}.
\newblock
  \bibinfo{howpublished}{\url{https://www.buzzfeed.com/jimwaterson/manchester-arena-fake-news}}.
\newblock
\newblock
\shownote{(Accessed on 12/02/2018).}


\bibitem[\protect\citeauthoryear{Wu and Liu}{Wu and Liu}{2018}]%
        {wu2018tracing}
\bibfield{author}{\bibinfo{person}{Liang Wu} {and} \bibinfo{person}{Huan Liu}.}
  \bibinfo{year}{2018}\natexlab{}.
\newblock \showarticletitle{Tracing fake-news footprints: Characterizing social
  media messages by how they propagate}. In
  \bibinfo{booktitle}{\emph{Proceedings of the Eleventh ACM International
  Conference on Web Search and Data Mining}}. ACM, \bibinfo{pages}{637--645}.
\newblock


\bibitem[\protect\citeauthoryear{Yang, Liu, Yu, and Yang}{Yang
  et~al\mbox{.}}{2012}]%
        {yang2012automatic}
\bibfield{author}{\bibinfo{person}{Fan Yang}, \bibinfo{person}{Yang Liu},
  \bibinfo{person}{Xiaohui Yu}, {and} \bibinfo{person}{Min Yang}.}
  \bibinfo{year}{2012}\natexlab{}.
\newblock \showarticletitle{Automatic detection of rumor on Sina Weibo}. In
  \bibinfo{booktitle}{\emph{Proceedings of the ACM SIGKDD Workshop on Mining
  Data Semantics}}. ACM, \bibinfo{pages}{13}.
\newblock


\bibitem[\protect\citeauthoryear{Zuckerman}{Zuckerman}{[n. d.]}]%
        {Stopsayi1:online}
\bibfield{author}{\bibinfo{person}{Ethan Zuckerman}.} \bibinfo{year}{[n.
  d.]}\natexlab{}.
\newblock \bibinfo{title}{Stop saying “fake news”. It’s not helping. |
  … My heart’s in Accra}.
\newblock
  \bibinfo{howpublished}{\url{http://www.ethanzuckerman.com/blog/2017/01/30/stop-saying-fake-news-its-not-helping/}}.
\newblock
\newblock
\shownote{(Accessed on 12/01/2018).}


\bibitem[\protect\citeauthoryear{Zuiderveen~Borgesius, Trilling, Moeller,
  Bod{\'o}, de~Vreese, and Helberger}{Zuiderveen~Borgesius
  et~al\mbox{.}}{2016}]%
        {zuiderveen2016should}
\bibfield{author}{\bibinfo{person}{Frederik Zuiderveen~Borgesius},
  \bibinfo{person}{Damian Trilling}, \bibinfo{person}{Judith Moeller},
  \bibinfo{person}{Bal{\'a}zs Bod{\'o}}, \bibinfo{person}{Claes~H de Vreese},
  {and} \bibinfo{person}{Natali Helberger}.} \bibinfo{year}{2016}\natexlab{}.
\newblock \showarticletitle{Should we worry about filter bubbles?}
\newblock  (\bibinfo{year}{2016}).
\newblock


\end{thebibliography}

\end{document}